\documentclass[preprint,review,3p,times,sort&compress]{elsarticle}

\usepackage{graphics}
\usepackage{graphicx}
\usepackage{subfigure}
\usepackage[numbers]{natbib}

\usepackage{psfrag}
\usepackage{amssymb}
\usepackage{amsmath}
\usepackage{mathrsfs}

\usepackage{amsthm}
\usepackage{xcolor}
\usepackage{rotating}
\usepackage{lscape}
\usepackage{pdflscape}
\usepackage{booktabs}
\usepackage{multirow}
\usepackage{hyperref}

\usepackage{lineno}
\usepackage{float}
\usepackage{epstopdf}
\usepackage{changes}
\usepackage{lipsum}

\colorlet{Changes@Color}{red}

\usepackage{hyperref}

\usepackage{ragged2e}

\begin{document}

\begin{frontmatter}




\title{Theoretical analysis of sound propagation and entropy generation across a distributed steady heat source }
%
\author[af1]{Jiaqi~Nan\corref{cor1}}
\ead{nan\_jiaqi@buaa.edu.cn}
\author[af1,af2]{Jingxuan~Li\corref{cor2}}
\ead{jingxuanli@buaa.edu.cn}
\author[af3]{Aimee S.~Morgans\corref{cor3}}
\ead{a.morgans@imperial.ac.uk}
\author[af1,af2]{Lizi~Qin\corref{cor5}}
\ead{qlz@buaa.edu.cn}
\author[af1,af2]{Lijun~Yang\corref{cor4}}
\ead{yanglijun@buaa.edu.cn}
\address[af1]{School of Astronautics, Beihang University, Beijing, China.}
\address[af2]{Aircraft and Propulsion Laboratory, Ningbo Institute of Technology, Beihang University, Ningbo, China.}
\address[af3]{Department of Mechanical Engineering, Imperial College London, London, UK.}
\cortext[cor4]{Corresponding author.}

%
%

\begin{abstract}
Acoustic and entropy waves interacting in a duct with a steady heat source and mean flow are analysed using an asymptotic expansion (AE) for low frequencies. The analytical AE solutions are obtained by taking advantage of flow invariants and applying a multi-step strategy. The proposed solutions provide first-order corrections to the compact model in the form of integrals of mean flow variables.   
An eigenvalue system is then built to predict the thermoacoustic modes of a duct containing a distributed heat source or sink. Predictions from the AE solutions agree well with the numerical results of the linearised Euler equations for both frequencies and growth rates, as long as the low-frequency condition is satisfied.
The AE solutions are able to accurately reconstruct the acoustic and entropy waves and correct the significant errors in the predicted entropy wave associated with the compact model. The analysis illustrates that the thermoacoustic system needs to account for the entropy wave generated by the interaction of acoustic wave and the distributed steady heat source, especially when density- or entropy-dependent boundary conditions are prescribed at the duct ends.   
Furthermore, a combination of the AE method and the modified WKB approximation method is discussed for a cooling case. The AE solutions remedy the disadvantage of the WKB solution in the low and very low-frequency domain and facilitate full-frequency theoretical analyses of sound propagation and entropy generation in inhomogeneous duct flow fields.

\end{abstract}
\end{frontmatter}
\section{Introduction}
\label{sec:intro}
Combustion systems in aero-engines, rockets and land-based gas turbines have a propensity to thermoacoustic instabilities, caused by positive feedback between acoustic perturbations and the unsteady heat release \citep{Yang_book_1995,Lieuwen_book_2012_UCP,Rienstra_Book_2016}. The resulting excessive pressure oscillations can cause component damage and even disastrous failure of the whole engine. Therefore, prediction and mitigation of thermoacoustic instabilities are major challenges during the design and testing of real engines.

Significant advances have been made in numerical and theoretical studies associated with the experimental demonstration of thermoacoustic instabilities \citep{Poinsot_Book_2005,Dowling_PCI_2015,Poinsot_PCI_2017}. Large Eddy Simulation (LES) is possible for full numerical prediction, but remains expensive and time-assuming for capturing the thermoacoustic properties of the whole system, because of the large disparity in the spatial-temporal scales of the multiphysics processes present in real industrial engines \citep{Gicquel_PECS_2012,Urbano_CNF_2016,Li_CNF_2017}. An alternative is to decouple the feedback of the acoustic system and the flame response and apply theoretical strategies for the former, and CFD simulations \citep{Han_CNF_2015,Li_CNF_2017} or experiments \citep{Schuller_CNF_2003,Noiray_JFM_2008,Palies_CNF_2011} for  the latter  . 
The overall system is separated geometrically into a sequence of connected modules (the flame being one of these), each with different acoustic properties. The acoustic field is then theoretically reconstructed by combining analytical solutions with matching conditions and end boundary conditions. Therefore, exact analytical solutions for each module enable the degrees of freedom of such network models \citep{Stow_ASME_2004,Xia_ASME_2017,Bonciolini_CNF_2021,YangD_JSV_2019_LONM} or Helmholtz equations \citep{Nicoud_AIAA_2007_AVBP,Silva_CNF_2013,Laera_JEGTP_2017,Laurent_JCP_2021} to be minimised.

Combustion systems often exhibit spatially non-uniform temperature distributions and varying mean flows due to the diffusion of combustion products and forced cooling \citep{Lieuwen_book_2012_UCP,Rienstra_Book_2016}. It is well known that the entropy wave generated by sound travelling through inhomogeneous regions \citep{Nicoud_CNF_2005,Wang_CNF_2019_entropy}  can be convected by the mean flow and reflected or transmitted by the boundaries \citep{Hield_AIAA_2008_choked, Weilenmann_Noiray_JSV_2021_choked}. Much attention has been paid to the distributed properties of unsteady heat sources \citep{Lieuwen_JSV_2000,Subramanian_CTM_2015,Laera_JEGTP_2017}. However, even when the heat source is purely steady, steady temperature gradients and mean flow acceleration mean that acoustic waves can be amplified or attenuated by interaction with entropy waves \citep{Karimi_JFM_2008,Karimi_JSV_2010,Bauerheim_CNF_2015, Laera_AIAACon_2017,Yeddula_JSV_2021}. This acoustic-entropy coupling consequently plays a crucial role in determining the frequencies and growth rates of thermoacoustic modes in the presence of temperature gradients \citep{Li_IJSCD_2017} and mean flows \citep{Nicoud_IJSCD_2009,Motheau_JSV_2014}. However, the effects of the mean temperature gradients associated with steady heat communication and the interaction of the acoustic and entropy waves is not appropriately quantified in the network models, these often being simplified by assuming no-flow or that the steady heat source is infinitely thin in spatial extent.
An accurate analytical solution which fully accounts for the acoustic-entropy interaction would be a technical improvement for low-order network models, enabling them to more accurately capture the thermoacoustic properties of real engines with inhomogeneous flow fields.

Extensive work on analytical solutions for sound propagation in ducts with temperature gradients or mean flows has been conducted \citep{Sujith_JSV_1995,Peat_JSV_1988,Cummings_JSV_1977,Li_JSV_2017a}. However, few studies fully account for the entropy wave. This is because its presence couples the governing partial differential equations (PDEs), which complicates their analytical solution in the presence of both large temperature gradients and non-zero mean flow effects. When decoupling or solving the PDEs, the resulting analytical solutions typically require trade-offs and balances between the frequencies over which they apply, the incoming flow Mach numbers they account for and the form of the temperature distribution.

At low frequencies, it is often assumed that the length of the distributed heat transferring region is much smaller than the acoustic wavelength. Then, the heat transferring region can be approximated as a spatial discontinuity which connects the upstream and downstream regions via matching conditions. The simplicity of this compact model approach makes it an attractive way of accounting for mean heat source temperature changes or the abrupt flow area variations into the framework of low order network models \citep{Dowling_JSV_1995,Morgans_CNF_2007,Schuller_JFM_2020}. It was also the approach applied to varying flows through choked nozzles in the absence of heat communication \citep{Marble_JSV_1977}.
In practice, flames and other regions of distributed flow properties often possess non-zero length compared to the acoustic wavelengths, even at low frequencies \citep{Stow_JFM_2002,Duran_JFM_2013,Li_CNF_2017}. As frequency increases, the compact assumption in which changes occur across an infinitely thin sheet give rise to significant errors; they further are unable to account for the important mechanism by which interaction of acoustic waves and the distributed mean temperature region give rise to acoustic-entropy coupling in the presence of mean flow \citep{Dowling_JSV_1995}. 

One strategy is to directly divide the distributed region into multiple subregions with constant or approximately linearly varying flow properties. Known analytical expressions can be applied in each subregion and subregions connected via matching conditions \citep{Moase_JFM_2007,Li_CNF_2015,ZhuSS_AST_2021_AM,Heilmann_JEGTP_2021}.
Another theoretical strategy to analyse the energy transport of an acoustic wave propagating through complicated flow distributions is using perturbation expansions \citep{Myers_JFM_1991,Nicoud_AIAA_2007_AVBP}. The thermoacoustic system is divided into subsystems of multiple orders; a higher-order system often corresponds to a more accurate correction which eventually approaches the exact solution. Frequency corrections to compact models, which used first-order asymptotic expansions in frequency to derive `effective lengths', were developed for adiabatic choked nozzles \citep{Stow_JFM_2002,Goh_JSV_2011}. 
However, for systems containing both complex mean temperature gradients and mean flows, perturbation variables have to be obtained by numerically resolving the subsystems of each order in turn. This numerical procedure makes it difficult to embed results into network models. Therefore, direct analytical solutions would offer substantial benefits for including the influence of distributed steady heat sources and the resulting acoustic-entropy interactions in the predictions of thermoacoustic instabilities.  

This paper presents an asymptotic analysis for low frequencies of the linearised Euler equations (LEEs) rewritten in terms of flow invariants. In Section \ref{sec:AE_method}, the analytical AE solutions for the acoustic and entropy waves across a distributed heat transferring region are derived in the form of a first-order frequency correction; they are further extended for higher frequencies via a multi-step strategy. For ducts subject to a passive flame and an incoming mean flow, the eigenvalue system is built using the proposed AE solutions. For a duct with an acoustically closed inlet and choked outlet, the acoustic and entropy waves are reconstructed and compared with those calculated using numerical solution of the LEEs and the compact model in Section \ref{sec:heating}. The combination of the AE solutions and the WKB solution is discussed in Section \ref{sec:combination} to achieve an accurate prediction across the full frequency range. Conclusions are drawn in Section \ref{sec:conclusion}. 

\section{Analytical transfer matrix of sound through a distributed steady heat source}
\label{sec:AE_method}
\begin{figure}[!ht]
\centering
\includegraphics[height=4cm]{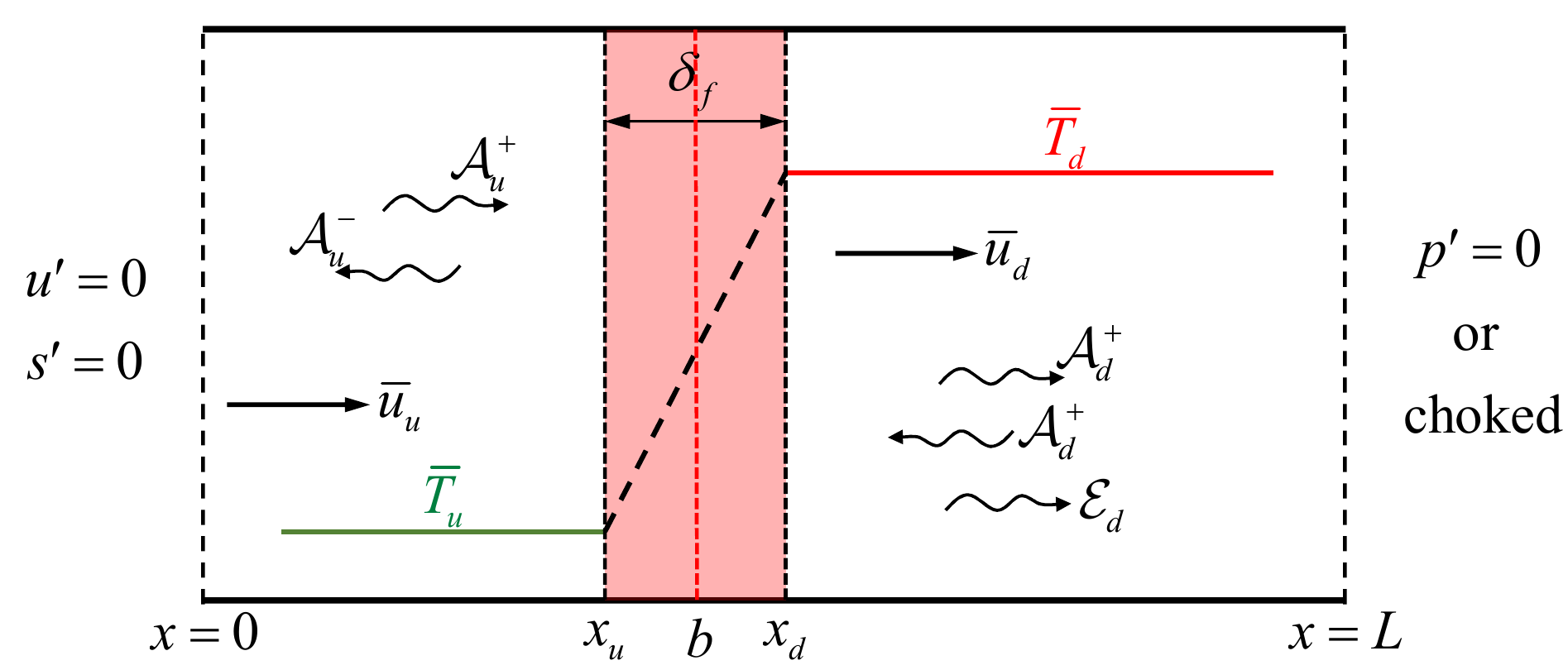}
\caption{Sketch of the one-dimensional duct with mean flow. The duct flow is subject to a region with mean temperature distribution extending over the shaded region. }
\label{Fig:Diagram}
\end{figure}

Consider the flow of inviscid perfect gas through a one-dimensional duct with a varying temperature region resulting from a heat source or forced cooling, see e.g. Fig.~\ref{Fig:Diagram} for a temperature raising system. 
The incoming mean flow encounters the distributed temperature region centred about $b = (x_d + x_u)/2$ and with thickness $\delta_f = x_d - x_u$. It is assumed that there is no entropy perturbation upstream the heat source or sink and the flow on both sides of it have constant mean flow properties unless otherwise stated. 
With the inlet acoustically closed, both the open-end ($p'=0$) and choked boundary conditions are applied to the duct outlet. The former is a pure acoustic condition. And the latter enables the acoustic wave to couple with the entropy wave convected by the mean flow and is representative of the outlets of many real combustion chambers. 
$\mathcal{A}^+$, $\mathcal{A}^-$ and $\mathcal{E}$ denote the amplitude of acoustic plane waves propagating in either direction and the entropy wave. The subscripts `$u$' and `$d$' represent steady flow variables upstream and downstream of the heat region.

\subsection{Dimensionless LEEs using an invariant vector across the distributed region}
\label{subsec:LEEs}
The one-dimensional Euler equations of mass, momentum and energy and the equation of state are applied to the spatially varying temperature region, $x\in[x_u,x_d]$, in the presence of an upstream mean flow and incident acoustic waves. These are written as:
\begin{equation}
\label{eq:Euler_mass}
\frac{\partial }{\partial t} (\rho) + \frac{\partial}{\partial x} (\rho u)=0\,,
\end{equation} 

\begin{equation}
\label{eq:Euler_momentum}
\frac{\partial}{\partial t}(\rho u) + \frac{\partial}{\partial x} (p + \rho u^2)=0\,,
\end{equation}

\begin{equation}
\label{eq:Euler_energy}
\frac{\partial}{\partial t}(\frac{p}{\gamma-1} + \frac{1}{2}\rho u^2) + \frac{\partial}{\partial x} \Big(\left[\frac{\gamma p}{\gamma-1} + \frac{1}{2}\rho u^2\right] u \Big)=\dot{q}\,,
\end{equation}

\begin{equation}
\label{eq:perfect_gas}
p=\rho R_g T\,,
\end{equation}
where $\rho$, $u$, $p$ and $T$ are density, velocity, pressure and temperature respectively and $\dot{q}$ represents the rate of heat transfer per unit volume. It should be noted that only the steady heat transfer is considered, with no heat perturbation, $\dot{q}^\prime=0$, assumed in the distributed region $x\in[x_u,x_d]$. Eqs.~\eqref{eq:Euler_mass}-\eqref{eq:perfect_gas} remain valid for a region with $\dot{q}\equiv0$. The specific heat ratio $\gamma$ and the universal gas constant $R_g$ are assumed constant along the duct.

In a linear system, each flow variable can be divided into a time-average steady component (denoted by $\bar{()}$) and a sufficiently small varying fluctuating component (denoted by $()^\prime=\hat{()}(x) e^{\mathrm{i}\omega t}$ ), where $\omega=2 \pi f -\mathrm{i}Gr$ is the complex angular frequency. $f$ and $Gr$ are the frequency and growth rate of each thermoacoustic mode. A positive value of $Gr$ indicates a growing wave.

For steady flow variables, the conserved equations can be expressed as:
\begin{equation}
 \label{eq:mean_m_governed}
\frac{\mathrm{d}}{\mathrm{d} x} (\bar{\rho} \bar{u})=0 ,
\end{equation}

\begin{equation}
 \label{eq:mean_f_governed}
\frac{\mathrm{d}}{\mathrm{d} x} (\bar{p} + \bar{\rho} \bar{u}^2)=0,
\end{equation}

\begin{equation}
 \label{eq:mean_e_governed}
\frac{\mathrm{d}}{\mathrm{d} x} \Big(\left[\frac{\gamma \bar{p}}{\gamma-1} + \frac{1}{2}\bar{\rho} \bar{u}^2\right] \bar{u} \Big)=\bar{\dot{q}}.
\end{equation}

From Eqs.~\eqref{eq:mean_m_governed} and \eqref{eq:mean_f_governed}, the steady mass and momentum flow rates per unit area remain constant along the inhomogeneous region $x\in [x_u, x_d]$, expressed as:
\begin{equation}
\label{eq:const_mass}
\bar{\rho}(x) \bar{u}(x) = \bar{\rho}_u \bar{u}_u = \bar{\dot{m}}
\end{equation}

\begin{equation}
\label{eq:const_momentum}
\bar{p}(x) + \bar{\rho}(x) {\bar{u}}^2(x) = \bar{p}_u + \bar{\rho}_u \bar{u}_u^2 = \bar{f}
\end{equation} 

Substituting the constant $\bar{\dot{m}}$ and the equation of state into momentum relation yields a quadratic equation for the mean velocity:
\begin{equation}
\bar{\dot{m}} {\bar{u}}^2(x)- \bar{f} \bar{u}(x) + \frac{\bar{T}(x)}{\bar{T}_u} \bar{u}_u \bar{p}_u =0
\end{equation}
For a given mean temperature distribution $\bar{T}(x)$, the mean velocity throughout the distributed region is obtained. The mean density $\bar{\rho}$ and pressure $\bar{p}$ are subsequently deduced from constant $\bar{\dot{m}}$ and $\bar{f}$, respectively.

The rate of steady heat transfer per unit area corresponding to $\bar{T}(x)$ is further reconstructed by integrating both sides of Eq.~\eqref{eq:mean_e_governed}, expressed as:
\begin{equation}
\bar{\dot{Q}}(x) = \int_{x_u}^{x} \bar{\dot{q}}(x) \mathrm{d} x=\bar{\rho}_u \bar{u}_u \left[ c_p \left(\bar{T}(x)-\bar{T}_u\right) + \frac{\bar{u}^2(x)-\bar{u}_u^2}{2}  \right]
\end{equation}
where $c_p= \gamma R_g / (\gamma-1)$ is the heat capacity at constant pressure. 

Therefore, the rate of steady heat transfer consistent with maintaining the mean temperature gradient and the corresponding varying mean flow field in the distributed region are reproduced by given upstream mean flow parameters and a prescribed temperature profile $\bar{T}(x)$.

Substituting both mean flow variables and the temporal periodic perturbations into Eqs.~\eqref{eq:Euler_mass}-\eqref{eq:Euler_energy} yields linearised Euler equations (LEEs):
\begin{equation}
\mathrm{i} \omega \hat{\rho} +\frac{\mathrm{d}}{\mathrm{d} x} ( \bar{\rho} \hat{u} + \hat{\rho} \bar{u} )= 0\,,
\end{equation}

\begin{equation}
\mathrm{i} \omega ( \bar{\rho} \hat{u} + \hat{\rho} \bar{u} ) + \frac{\mathrm{d}}{\mathrm{d} x} (\hat{p} + 2 \bar{\rho} \bar{u} \hat{u} + \hat{\rho} \bar{u}^2 )= 0\,,
\end{equation}

\begin{equation}
\mathrm{i} \omega (\frac{\hat{p}}{\gamma-1}+ \bar{\rho} \bar{u} \hat{u} +   \frac{\hat{\rho} \bar{u}^2}{2} ) + \frac{\mathrm{d}}{\mathrm{d} x} ( \frac{\gamma \hat{p}}{\gamma-1}\bar{u}  + \frac{\hat{\rho} \bar{u}^3}{2}  + \frac{ 3 \bar{\rho} \bar{u}^2}{2} \hat{u} + \frac{\gamma \bar{p}}{\gamma-1} \hat{u} )
= \hat{\dot{q}}\,.
\end{equation}

All flow variables in LEEs are normalised by the upstream mean parameters:
\begin{equation}
\label{eq:reduced_variables}
\boldsymbol{P}=\frac{p}{\bar{\rho}_u \bar{c}_u^2} , \quad \boldsymbol{U}=\frac{u}{\bar{c}_u} ,  \quad  \boldsymbol{\rho}=\frac{\rho}{\bar{\rho}_u} , \quad  X=\frac{x}{L}, \quad \Omega = \frac{\omega L}{\bar{c}_u}, \quad \bar{\boldsymbol{c}}=\frac{\bar{c}}{\bar{c}_u}, \quad  M = \frac{\bar{\boldsymbol{U}}}{\bar{\boldsymbol{c}}}, 
\end{equation}
where $L$ is the duct length and $X$ is the normalised axial position within the finite distributed region. $\bar{c}=(\gamma R_g \bar{T})^{1/2}$ is the speed of sound and $M$ is the local flow Mach number.

The dimensionless linearised Euler equations (LEEs) can then be written as follows with the frequency-dependent terms all moved to the right hand side: 
\begin{equation}
\label{eq:mass_LEEs}
\frac{\mathrm{d}}{\mathrm{d} X} ( \bar{\boldsymbol{\rho}} \hat{\boldsymbol{U}} + \hat{\boldsymbol{\rho}} \bar{\boldsymbol{U}} )= - \mathrm{i} \Omega \hat{\boldsymbol{\rho}}\,,
\end{equation}

\begin{equation}
\label{eq:momentum_LEEs}
\frac{\mathrm{d}}{\mathrm{d} X} (\hat{\boldsymbol{P}} + 2 \bar{\boldsymbol{\rho}} \bar{\boldsymbol{U}} \hat{\boldsymbol{U}} + \hat{\boldsymbol{\rho}} \bar{\boldsymbol{U}}^2 )= - \mathrm{i} \Omega ( \bar{\boldsymbol{\rho}} \hat{\boldsymbol{U}} + \hat{\boldsymbol{\rho}} \bar{\boldsymbol{U}})\,,
\end{equation}

\begin{equation}
\label{eq:energy_LEEs}
\frac{\mathrm{d}}{\mathrm{d} X} ( \frac{\gamma \hat{\boldsymbol{P}}}{\gamma-1}\bar{\boldsymbol{U}}  + \frac{\hat{\boldsymbol{\rho}} \bar{\boldsymbol{U}}^3}{2}  + \frac{ 3 \bar{\boldsymbol{\rho}} \bar{\boldsymbol{U}}^2}{2} \hat{\boldsymbol{U}} + \frac{\gamma \bar{\boldsymbol{P}}}{\gamma-1} \hat{\boldsymbol{U}} )
= - \mathrm{i} \Omega (\frac{\hat{\boldsymbol{P}}}{\gamma-1}+ \bar{\boldsymbol{\rho}} \bar{\boldsymbol{U}} \hat{\boldsymbol{U}} +   \frac{\hat{\boldsymbol{\rho}}\bar{\boldsymbol{U}}^2}{2} )\,.
\end{equation}
As mentioned previously, this work assumes that there is no heat perturbation in the steady heat source region, $\hat{\dot{q}}=0$.

The left hand sides of the equations allow an invariant vector, $\boldsymbol{I}$, to be defined for the flow travelling through the duct; this results from transforming the acoustic perturbation vector via a transfer matrix $\mathcal{M}^{p2I}$ at any position.
\begin{equation}
\label{eq:M_p2I}
\boldsymbol{I}=
\mathcal{M}^{p2I}
\begin{bmatrix}     
\hat{\boldsymbol{P}}   \vspace{0.2cm}\\     
\hat{\boldsymbol{\rho}}  \vspace{0.2cm}\\ 
\hat{\boldsymbol{U}}   \\
\end{bmatrix}
=
\begin{bmatrix}     
0     &\bar{\boldsymbol{U}}   & \bar{\boldsymbol{\rho}}   \vspace{0.2cm}\\     
1     &\bar{\boldsymbol{U}}^2  & 2 \bar{\boldsymbol{\rho}} \bar{\boldsymbol{U}}  \vspace{0.2cm}\\ 
\dfrac{\gamma \bar{\boldsymbol{U}} }{\gamma-1}  &\dfrac{1}{2}\bar{\boldsymbol{U}}^3  
&\dfrac{3}{2} \bar{\boldsymbol{\rho}}\bar{\boldsymbol{U}}^2 + \dfrac{\gamma \bar{\boldsymbol{P}} }{\gamma-1}   \\
\end{bmatrix}   
\begin{bmatrix}     
\hat{\boldsymbol{P}}   \vspace{0.2cm}\\     
\hat{\boldsymbol{\rho}}  \vspace{0.2cm}\\ 
\hat{\boldsymbol{U}}   \\
\end{bmatrix}.
\end{equation}
Noted that the subscripts 'u' and 'd' in the $\mathcal{M}^{p2I}$ indicate the matrix at $X=X_u$ and $X=X_d$ in the following derivation. It follows that $\mathcal{M}^{p2I}$ is same for $X\in[0,X_u]$ and $X\in[X_d,L]$ due to the constant flow properties in both regions upstream and downstream of the inhomogeneous region.

Rewriting Eqs.~\eqref{eq:mass_LEEs}-\eqref{eq:energy_LEEs} using the invariant, one obtains a set of ordinary differential equations (ODEs) with spatially varying coefficients:
\begin{equation}
\label{eq:wave_eq}
\frac{\mathrm{d} \boldsymbol{I}}{\mathrm{d} X}  =   \boldsymbol{A}(X)\boldsymbol{I}
\end{equation}
where $\boldsymbol{A}$  is a complex coefficient matrix with terms dependent on the axial position $X$,
\begin{equation}
\boldsymbol{A}(X)=\mathcal{M}_{1} (\mathcal{M}^{p2I})^{-1}=
- \mathrm{i} \Omega 
\begin{bmatrix}     
0 &1 &0 \vspace{0.2cm}\\     
0 &\bar{\boldsymbol{U}} &\bar{\boldsymbol{\rho}} \vspace{0.2cm}\\ 
\dfrac{1}{\gamma-1} &\dfrac{1}{2}\bar{\boldsymbol{U}}^2  &\bar{\boldsymbol{\rho}} \bar{\boldsymbol{U}}  \\
\end{bmatrix}
\begin{bmatrix}     
0     &\bar{\boldsymbol{U}}   & \bar{\boldsymbol{\rho}}   \vspace{0.2cm}\\     
1     &\bar{\boldsymbol{U}}^2  & 2 \bar{\boldsymbol{\rho}} \bar{\boldsymbol{U}}  \vspace{0.2cm}\\ 
\dfrac{\gamma \bar{\boldsymbol{U}} }{\gamma-1}  &\dfrac{1}{2}\bar{\boldsymbol{U}}^3  
&\dfrac{3}{2} \bar{\boldsymbol{\rho}}\bar{\boldsymbol{U}}^2 + \dfrac{\gamma \bar{\boldsymbol{P}} }{\gamma-1}   \\
\end{bmatrix}^{-1}\,,
\end{equation}
and $\mathcal{M}^{p2I}$ can be inverted when its determinant is non-zero,
\begin{equation}
\det(\mathcal{M}^{p2I})=\frac{\gamma \bar{\boldsymbol{P}}}{\gamma-1}\bar{\boldsymbol{U}}(M^2-1)\neq 0\,.
\end{equation}

Eq.~\eqref{eq:wave_eq} holds when the mean velocity is non-zero, $\bar{\boldsymbol{U}}\neq0$, and the bulk flow is not choked along the duct, $M\neq1$. The case of no mean duct flow will be separately discussed in Sec.~\ref{subsec:no mean flow}. Note that accounting for choked flow would be important in many nozzles and is not discussed in this paper.

Hence the problem of incident acoustic waves propagating through a distributed region of steady heat source is transformed into an initial value problem of linear ODEs with a spatially varying coefficient matrix. The next subsection will present why the invariant vector makes it possible to obtain analytical expressions for the acoustic transmission and the entropy generation across the flow variation region.


\subsection{Asymptotic analysis for low frequencies}
\label{subsec:AE_solution}
Analytical solutions of linear ODEs typically take the form of integration of the exponential function of the coefficient matrix. It has been shown that for the LEEs of interest in the present work, the resulting linear ODEs with their spatially varying coefficient matrix can be resolved using a high-order Magnus expansion (ME); this involves an exponential expression of nested operators \citep{Magnus_1954, Duran_JFM_2013}. The ME method thus often relies on recursive procedures or is embedded in a numerical scheme. Its expression also does not provide any physical insight into the sound transmission and entropy generation induced by the interaction of acoustic waves with the distributed steady heat source.

To obtain a direct expression, an asymptotic analysis for a small $\Omega$ is conducted here. The acoustic perturbations are written as 
\begin{equation}
\label{eq:AE_p}
\hat{\boldsymbol{P}}= \hat{\boldsymbol{P}}_0 + \mathrm{i} \Omega \hat{\boldsymbol{P}}_1 + O({\Omega}^2), \quad 
\hat{\boldsymbol{\rho}}= \hat{\boldsymbol{\rho}}_0 + \mathrm{i} \Omega \hat{\boldsymbol{\rho}}_1 + O({\Omega}^2), \quad 
\hat{\boldsymbol{U}}= \hat{\boldsymbol{U}}_0 + \mathrm{i} \Omega \hat{\boldsymbol{U}}_1 + O({\Omega}^2).
\end{equation}
The invariant vector then has the same expansion with the second-order term $O({\Omega}^2)$ neglected.
\begin{equation}
\label{eq:AE_I}
\boldsymbol{I}= \boldsymbol{I}_0 + \mathrm{i} \Omega \boldsymbol{I}_1 + O({\Omega}^2)
\end{equation}

Substituting Eq.~\eqref{eq:AE_I} into Eq.~\eqref{eq:wave_eq} leads to the zero-order and first-order systems:
\begin{equation}
\label{eq:0st_order_ODEs}
\frac{\mathrm{d} \boldsymbol{I}_0}{\mathrm{d} X}  = 0
\end{equation}
\begin{equation}
\label{eq:1st_order_ODEs}
\frac{\mathrm{d} \boldsymbol{I}_1}{\mathrm{d} X}  =  \boldsymbol{A}_1(X) \boldsymbol{I}_0
\end{equation}
where $\boldsymbol{A}_1(X)  =  \boldsymbol{A}(X)/\mathrm{i} \Omega $.

It is found that the zero-order invariant, $\boldsymbol{I}_0$ in Eq.~\eqref{eq:0st_order_ODEs}, is constant with axial position, $X$, and can be determined from the upstream mean flow variables using:
\begin{equation}
\label{eq:constant_I_0}
\boldsymbol{I}_0 \equiv \boldsymbol{I}_0(X_u) =
\mathcal{M}^{p2I}_u     
\begin{bmatrix}
\hat{\boldsymbol{P}}  & \hat{\boldsymbol{\rho}} &\hat{\boldsymbol{U}}
\end{bmatrix}^{\mathrm{T}}_u\,.
\end{equation}
This invariance of $\boldsymbol{I}_0$ enables the first-order term $\boldsymbol{I}_1$ in Eq.~\eqref{eq:1st_order_ODEs} to be derived by direct integration, element by element, as:
\begin{equation}
\label{eq:I_1}
\begin{split}
\boldsymbol{I}_1(X) &=\left[\int_{X_u}^{X}  \boldsymbol{A}_1(X)\mathrm{d}X \right] \boldsymbol{I}_0\\
&= \left( - \int_{X_u}^{X} \mathrm{d}X 
\begin{bmatrix}     
\dfrac{1}{\bar{\boldsymbol{U}}}+\dfrac{(1+\gamma) M^2}{2 (M^2-1) \bar{\boldsymbol{U}}} &\dfrac{-\gamma}{\bar{\boldsymbol{c}}^2 (M^2 - 1)}  & \dfrac{\gamma-1}{\bar{\boldsymbol{U}} \bar{\boldsymbol{c}}^2 (M^2-1)}\vspace{1em} \\
1  &0  &0\vspace{1em} \\
-\dfrac{\bar{\boldsymbol{U}} (\gamma-3) (2-M^2+\gamma M^2)}{4(\gamma-1) (M^2-1)}     &\dfrac{1}{\gamma-1} + \dfrac{(\gamma^2-3 \gamma)M^2}{2(\gamma-1)(M^2-1)}   &- \dfrac{\bar{\boldsymbol{U}}(\gamma-3)}{2 \bar{\boldsymbol{c}}^2 (M^2-1)}
\end{bmatrix} \right) \boldsymbol{I}_0\,,
\end{split}
\end{equation}
as long as $\mathcal{M}^{p2I}$ can be inverted ($\bar{\boldsymbol{U}}(X)\neq0$ and $M(X)\neq1$) in the distributed temperature region, $[X_u,X_d]$.

Consequently, the asymptotic expansion (AE) solution for the invariant vector is expressed as:
\begin{equation}
\label{eq:solution_I}
\boldsymbol{I}(X) = \boldsymbol{I}_0 + \mathrm{i}\Omega\boldsymbol{I}_1 
=\left[\boldsymbol{\mathrm{M}}_I\right]_{X_u}^{X} \boldsymbol{I}_0\,,
\end{equation}
where the transfer matrix for the invariant at $X\in [X_u,X_d]$ is given by
\begin{equation}
\label{eq:Trans_M}
\left[\boldsymbol{\mathrm{M}}_I\right]_{X_u}^{X}= \boldsymbol{E}+ \mathrm{i}\Omega\int_{X_u}^{X}  \boldsymbol{A}_1(X)\mathrm{d}X ~,
\end{equation}
and $\boldsymbol{E}$ represents the identity matrix.

It is worth noting that the same results could be derived for $\boldsymbol{\mathrm{M}}_I$ when the reference length $L$ in the  parameters $\Omega$ and $X$ are replaced with the distributed thickness $\delta_f$.   Therefore, the duct length $L$ is always used for the normalisation throughout this paper to make the dimensionless parameters of the temperature variation region consistent with those of the whole duct.

Normalised acoustic perturbations are then obtained via transformation of the invariant:
\begin{equation}
\label{eq:solution_acoustic_pertubations}
\begin{bmatrix}     
\hat{\boldsymbol{P}}  &\hat{\boldsymbol{\rho}} &\hat{\boldsymbol{U}} 
\end{bmatrix}^{\mathrm{T}} = (\mathcal{M}^{p2I})^{-1} \boldsymbol{I}\,.
\end{equation}
The entropy perturbation generated by the interaction of incident acoustic waves and mean temperature gradient is calculated from the entropy relation, $\hat{s}/c_p = \hat{p}/\gamma \bar{p} - \hat{\rho}/\bar{\rho}$, and is given by:
\begin{equation}
\hat{\boldsymbol{\sigma}} =\frac{\hat{\boldsymbol{P}}}{\gamma \bar{\boldsymbol{P}}}-\frac{\hat{\boldsymbol{\rho}}}{\bar{\boldsymbol{\rho}}}\,,
\end{equation}
where $c_p$ is the heat capacity at constant pressure.
 
Finally, the acoustic transfer matrix from $X=X_u$ to $X=X_d$ is obtained:
\begin{equation}
\label{eq:FTF}
\begin{bmatrix}     
\hat{\boldsymbol{P}}  \vspace{1em} \\
\hat{\boldsymbol{\rho}}  \vspace{1em} \\
\hat{\boldsymbol{U}}
\end{bmatrix}_d
=\left[\boldsymbol{\mathrm{T}}\right]_{u}^{d}
\begin{bmatrix}     
\hat{\boldsymbol{P}}  \vspace{1em} \\
\hat{\boldsymbol{\rho}} \vspace{1em} \\
\hat{\boldsymbol{U}}
\end{bmatrix}_u
 = (\mathcal{M}^{p2I}_d)^{-1}\left[\boldsymbol{\mathrm{M}}_I\right]_{X_u}^{X_d} \mathcal{M}^{p2I} _u
 \begin{bmatrix}     
\hat{\boldsymbol{P}}  \vspace{1em} \\
\hat{\boldsymbol{\rho}} \vspace{1em} \\
\hat{\boldsymbol{U}}
\end{bmatrix}_u\,.
\end{equation}%

It should be noted that $\boldsymbol{\mathrm{M}}_I$ is equal to the identity matrix $\boldsymbol{E}$ when the distributed region is infinitely thin ($X_d=X_u$); the acoustic transfer matrix for such a concentrated steady heat source is 
\begin{equation}
\label{eq:FTF_compact}
\left[\boldsymbol{\mathrm{T}}_c\right]_{u}^{d}
 = (\mathcal{M}^{p2I}_d)^{-1} \mathcal{M}^{p2I}_u\,.
\end{equation}%
%
\subsection{Acoustic transfer matrix for no mean flow}
\label{subsec:no mean flow}
For no mean flow or when the Mach numbers are very small ($M\approx0$),  $\mathcal{M}^{p2I}$ is effectively singular everywhere and cannot be inverted to obtain the acoustic transfer matrix. The case of no mean flow is therefore treated separately in this subsection.

According to the conservation of steady momentum, no mean flow yields constant mean pressure, $\gamma \bar{\boldsymbol{P}} = 1$.
With $\bar{\boldsymbol{U}}=0$, the dimensionless LEEs in Eqs.~\eqref{eq:mass_LEEs}-\eqref{eq:energy_LEEs} then simplify to
\begin{equation}
\label{eq:LEE_P_noflow}
\frac{\mathrm{d} \hat{\boldsymbol{P}}}{\mathrm{d} X} = -\mathrm{i} \Omega \bar{\boldsymbol{\rho}} \hat{\boldsymbol{U}}\,,
\end{equation}

\begin{equation}
\label{eq:LEE_U_noflow}
\frac{\mathrm{d} \hat{\boldsymbol{U}}}{\mathrm{d} X} = -\mathrm{i} \Omega  \hat{\boldsymbol{P}}\,.
\end{equation}
The acoustic perturbations are subsequently derived using the asymptotic expansion in Eq.~\eqref{eq:AE_p}, written as:
\begin{equation}
\label{eq:FTF_noFlow}
\begin{bmatrix}     
\hat{\boldsymbol{P}}(X)  \vspace{1em} \\
\hat{\boldsymbol{U}}(X)
\end{bmatrix}
 = \big[\boldsymbol{\mathrm{T}}_\mathrm{noFlow}\big]_{u}^{X}
 \begin{bmatrix}     
\hat{\boldsymbol{P}} \vspace{1em} \\
\hat{\boldsymbol{U}}
\end{bmatrix}_u
 = 
\begin{bmatrix}     
1 & -\mathrm{i} \Omega \int_{X_u}^{X}  \bar{\boldsymbol{\rho}} \mathrm{d} X \vspace{1em} \\
-\mathrm{i} \Omega (X-X_u) &1
\end{bmatrix}
\begin{bmatrix}     
\hat{\boldsymbol{P}} \vspace{1em} \\
\hat{\boldsymbol{U}}
\end{bmatrix}_u\,.
\end{equation}
The density perturbation is then given by $\hat{\boldsymbol{\rho}} = \hat{\boldsymbol{P}}/\bar{\boldsymbol{c}}^2$.

The acoustic transfer matrix for no mean flow can then be derived across the flow variation region, and is expressed as:
\begin{equation}
\label{eq:FTF_noFlow_ud}
\begin{bmatrix}     
\hat{\boldsymbol{P}}  &\hat{\boldsymbol{U}}
\end{bmatrix}^{\mathrm{T}}_d
 = \left[\boldsymbol{\mathrm{T}}_\mathrm{noFlow}\right]_{u}^{d}
\begin{bmatrix}     
\hat{\boldsymbol{P}} &\hat{\boldsymbol{U}}
\end{bmatrix}^{\mathrm{T}}_u\,.
\end{equation}
\subsection{Multi-step AE solutions for smaller Mach numbers and higher frequencies}
\label{subsec:AE_steps}
The proposed AE solutions are valid at low frequencies, when the second-order frequency term $O(\Omega^2)$ is negligible. It is expected that the AE solutions for low frequencies will have a limit of validity as $\Omega$ increases. Furthermore, the entropy wave, generated by the incident acoustic wave travelling through the distributed flow variation region, has a much smaller wavelength than the acoustic waves when the ratio $\Omega/M$ is relatively large. This smaller entropic wavelength makes it impractical to accurately predict the entropy disturbance using solution of the previous one-step integration over the finite-length region. 

The AE solutions using the invariant vector are therefore extended for smaller Mach numbers and higher frequencies via a multi-step strategy. The normalised interval $[X_u, X_d]$ is divided into $N_s$ subintervals $[X_{i-1},X_i]$, where $i=1,2,\cdots, N_s$. 
Transfer matrices for the invariant are calculated for each subinterval, expressed as:
\begin{equation}
\left[\boldsymbol{\mathrm{M}}_I\right]_{{i-1}}^{i}
= \left[\boldsymbol{E}+ \mathrm{i}\Omega \int_{X_{i-1}}^{X_i} \boldsymbol{A}_1(X)~ \mathrm{d}X \right]\,.
\end{equation}
By multiplying discrete transfer matrices in turn, the acoustic transfer matrix of the multi-step AE solution is obtained from $X=X_u$ to $X=X_d$, as:
\begin{equation}
\left[\boldsymbol{\mathrm{T}}\right]_{u}^{d}
 = (\mathcal{M}^{p2I}_d)^{-1} \prod\limits_{1}^{N_s} \left[\boldsymbol{\mathrm{M}}_I\right]_{{i-1}}^{i} \mathcal{M}^{p2I}_u\,.
\end{equation}

This method exploits the fact that the entropy perturbation can be exactly captured when the length of each subinterval $\Delta x_i=|X_i-X_{i-1}|$ is sufficiently smaller than the entropy wavelength. Two normalised ratios are now defined to quantify the frequency limit and step size for the multi-step AE solutions to predict the entropy and acoustic perturbations, respectively. These are given by: 

\begin{equation}
\label{eq:limits_AE_s}
\zeta_s=\frac{\Delta x_i}{\lambda_s} = \frac{\Delta x_{i}}{\delta_f} \frac{\Omega/2\pi}{M} \frac{\delta_f}{L}  \frac{\bar{c}_u}{\bar{c}} \ll 1\,,
\end{equation}

\begin{equation}
\label{eq:limits_AE_a}
\zeta_a=\frac{\Delta x_i}{\lambda_a} = M \cdot \zeta_s = \frac{\Delta x_{i}}{\delta_f} \frac{\Omega}{2\pi} \frac{\delta_f}{L}  \frac{\bar{c}_u}{\bar{c}}\,.
\end{equation}
where the subscripts `$s$' and `$a$' denote entropy and acoustic waves, respectively, and $\lambda$ is the corresponding wavelength.

The normalised entropy ratio, $\zeta_s$, for the discrete distributed region depends on $\Omega/M$. This demonstrates that the AE solutions for higher frequencies or smaller Mach numbers are of poor accuracy until small enough step sizes are applied. The same is true for larger distribution distances relative to the duct length, $\delta_f/L$. For the acoustic ratio, $\zeta_a$, the step size depends only on the frequency, $\Omega$, and  not the Mach number $M$. This will be discussed in detail in Section~\ref{sec:EigenFre_with_flame}. 

By defining a low-frequency condition, $\Omega_{\mathrm{LF}}$ through $\zeta_s = \Omega/\Omega_{\mathrm{LF}}$, the expression for this condition for the proposed multi-step AE solution is:

\begin{equation}
\label{eq:LFC_s_for_Omega}
\Omega \ll \Omega_{\mathrm{LF}} =  \frac{2 \pi L}{\bar{c}_u} \cdot \min{\{M, 1\}} \cdot \min_{i}{\left\{ \frac{\bar{c}}{\Delta x_i}  \right\}}\,,
\end{equation}

In the absence of an upstream mean flow, the critical value for the acoustic low-frequency condition is given by
\begin{equation}
\label{eq:LFC_a_for_noFlow}
\Omega \ll \Omega_{\mathrm{LF,a}} =\frac{2 \pi L}{\bar{c}_u} \cdot \min_{i}{\left\{ \frac{\bar{c}}{\Delta x_i}  \right\}}
\end{equation}
and $\zeta_a = \Omega/\Omega_{\mathrm{LF,a}}$.
$\Delta x = \delta_f$ is the case of the one-step AE method over the whole distributed region. The above expression suggests that the step size could be optimally chosen from the mean flow field and frequencies in order to accurately predict the acoustic and entropy waves through a duct with distributed steady heat source.

\section{Duct containing a finite-length mean temperature distribution}
\label{sec:heating}
\begin{figure}[!ht]
	\centering
	\subfigure
  		{
\includegraphics[width=7cm]{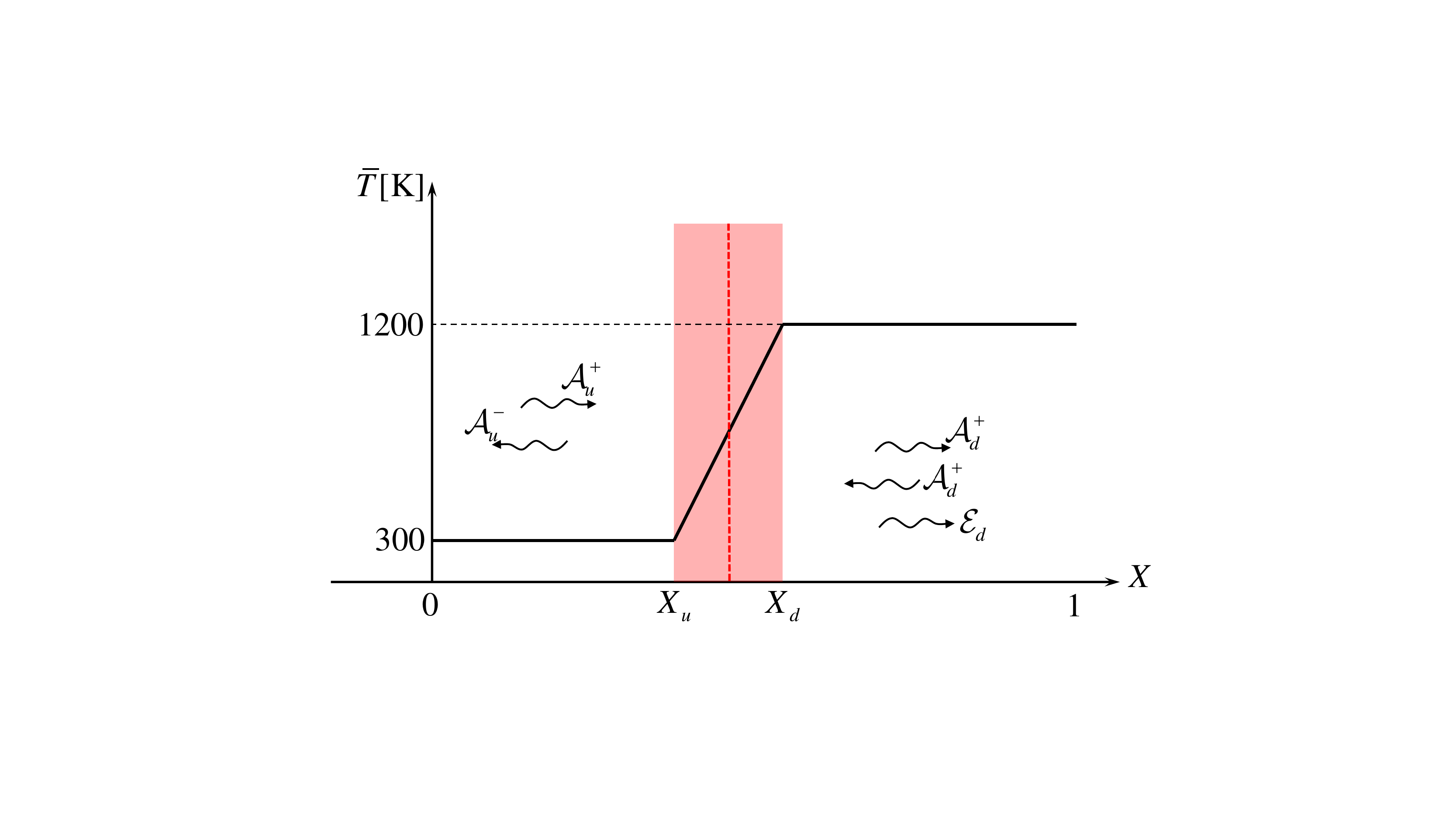}
  		}
  		\put (-210,105) {\normalsize$\displaystyle(a)$}
  		\vspace*{-0pt}
  		\hspace*{0pt}
  		\subfigure
  		  		{
\includegraphics[width=7cm]{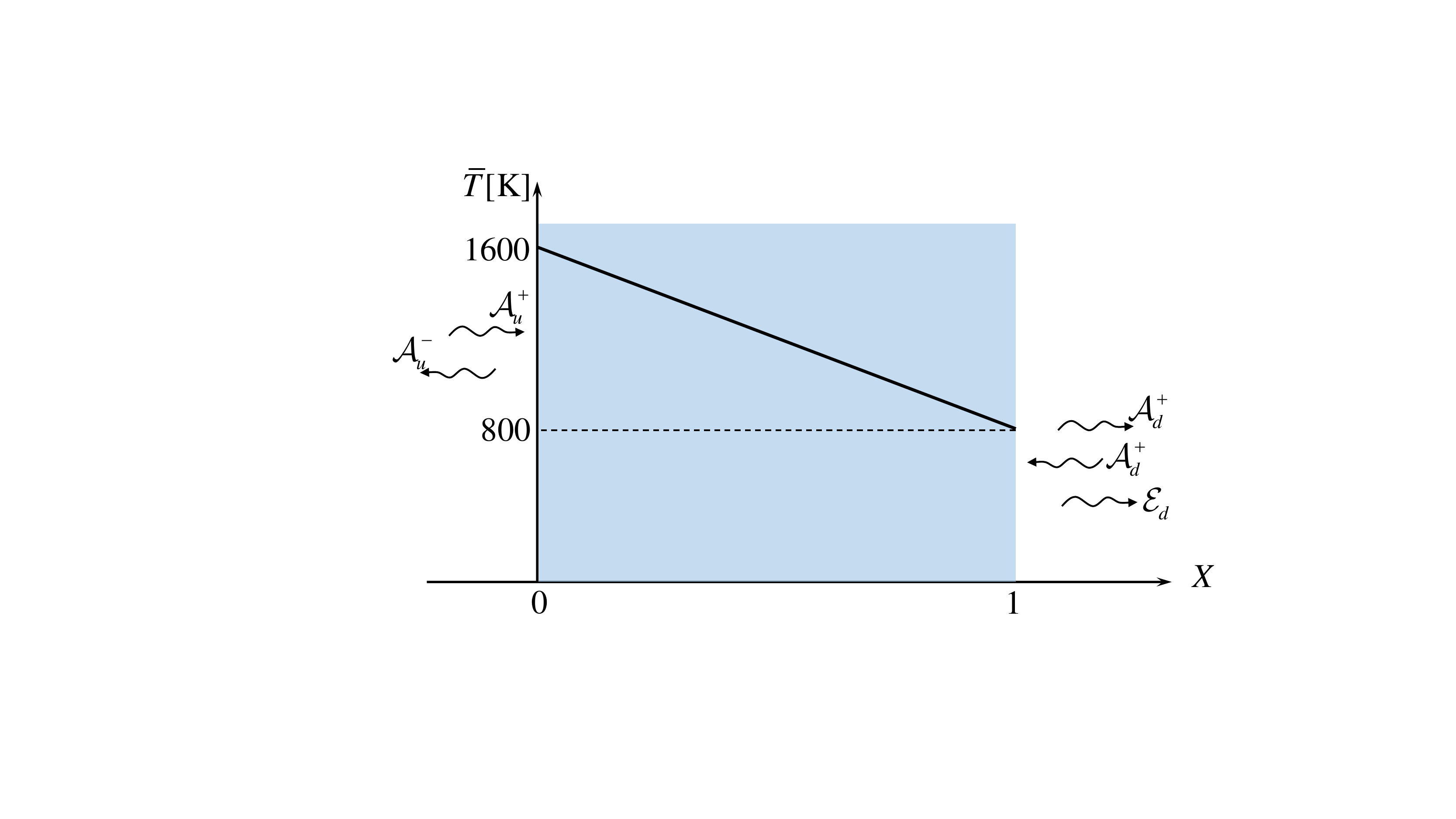}
  		}
  		\put (-210,105) {\normalsize$\displaystyle(b)$}
  		\vspace*{-0pt}
 	 \caption{Distributions of mean temperature $\bar{T}$ with the normalised axial position $X$ for (a) a passive flame and (b) a pure cooling case. The centre position of the passive flame is represented by the red dotted line (\textcolor{red}{$\mathbf{- -}$}).}
	 \label{Fig:Sketch_Cases}
	 \vspace*{00pt}
\end{figure}

The distributed heating region resulting from a passive (i.e. steady) flame or a cooling region is embedded in a straight duct with no entropy perturbation upstream considered, as sketched in Fig.~\ref{Fig:Sketch_Cases}. Note that $\gamma=1.4$ and $R_g = 287~\mathrm{J}\cdot \mathrm{K}^{-1} \cdot \mathrm{kg}^{-1}$.

The proposed AE solutions are implemented to predict the propagation and generation of acoustic and entropy waves across these mean temperature distributions. Four transfer functions are defined, from the normalised pressure perturbation at the duct inlet to the acoustic and entropic variables at any position, expressed as:
\begin{equation}
\label{eq:TF}
\mathcal{F}_p(X, \Omega) =  \frac{\hat{\boldsymbol{P}}}{\hat{\boldsymbol{P}}_0},~~~~
\mathcal{F}_{s}(X, \Omega) = \frac{\hat{\boldsymbol{\sigma}}}{\hat{\boldsymbol{P}}_0}.
\end{equation}

The duct length is taken to be fixed at $L=1$m unless otherwise specified. To quantify the differences between the transfer functions calculated by the AE solutions and the numerical results of LEEs (Eqs.~\eqref{eq:mass_LEEs}-\eqref{eq:energy_LEEs}), error coefficients are defined as

\begin{equation}
\epsilon_p (\mathcal{F}_{p,\mathrm{AE}},\mathcal{F}_{p,\mathrm{LEEs}}) = \left( \sum_i \Big|\mathcal{F}_{p,\mathrm{AE}}(X_i)-\mathcal{F}_{p,\mathrm{LEEs}}(X_i)\Big|^2 \Bigg/ \sum_i \Big|\mathcal{F}_{p,\mathrm{LEEs}}(X_i)\Big|^2\right)^{1/2}\,,
\end{equation}
\begin{equation}
\epsilon_s (\mathcal{F}_{s,\mathrm{AE}},\mathcal{F}_{s,\mathrm{LEEs}}) = \left( \sum_i \Big|\mathcal{F}_{s,\mathrm{AE}}(X_i)-\mathcal{F}_{s,\mathrm{LEEs}}(X_i)\Big|^2 \Bigg/ \sum_i \Big|\mathcal{F}_{s,\mathrm{LEEs}}(X_i)\Big|^2\right)^{1/2}\,.
\end{equation}

In Sec.~\ref{sec:EigenFre_with_flame}, the analytical AE solutions are validated by comparing them to the numerical results of both the LEEs and the Magnus expansion (ME). In Sec.~\ref{sec:EigenFre_with_flame_thickness}, the eigenvalue system for a duct with an acoustically closed inlet and choked outlet is then built using the different methods. The eigenfrequencies and growth rates of the thermoacoustic modes are calculated, with the effect of incoming flow Mach number evaluated in Sec.~\ref{sec:EigenFre_with_Ma}.

\subsection{Transfer functions for distributed steady heating region}
\label{sec:EigenFre_with_flame}
The acoustic transfer matrix of the proposed AE method is applied to the distributed steady heating region. The incoming mean flow is assumed to be heated such that the temperature increases linearly from $\bar{T}_u=300$K to $\bar{T}_d=1200$K over a finite axial distance $\delta_f=0.1$m, with the distributed region starting at $X=X_u$ and ending at $X=X_d$. A local ``flame'' coordinate, $x_f  \in [0, \delta_f]$, is used to characterise the axial variations of the transfer functions inside the distributed heat source, and is normalised by the flame thickness such that $X_f=x_f/\delta_f$.

The acoustic and entropy transfer functions are obtained for an upstream incoming acoustic wave with given frequency $f$; there are no entropy perturbation upstream of the heating region. The incident pressure and velocity disturbances are determined using $\hat{p}_u =  \bar{\rho}_u \bar{c}_u Z_u \hat{u}_u$ with an impedance of $Z_u=-1$ applied at $X=X_u$. The upstream density disturbance follows from the condition of zero entropy disturbance $\hat{\sigma}_u=0$ at $X=X_u$. The normalised acoustic perturbations upstream of the heating region can therefore be expressed as: 

\begin{equation}
\label{eq:BC_Xu}
\begin{bmatrix}     
\hat{\boldsymbol{P}}  &\hat{\boldsymbol{\rho}} &\hat{\boldsymbol{U}}  
\end{bmatrix}^{\mathrm{T}}_u
=
\begin{bmatrix}     
1  &1  & 1/Z_u
\end{bmatrix}^{\mathrm{T}}  \hat{\boldsymbol{P}}_u \,.
\end{equation} 

The analytical AE solutions then allow the acoustic field and transfer functions inside the heated region to be reconstructed utilising the one-step and multi-step strategies. The latter is used for larger $\Omega/M_u$, with uniform grids containing 1000 axial segments applied. The AE solutions are validated by comparing them to the numerical results of LEEs that are calculated via the second-order finite difference scheme on the same uniform grids as the multi-step AE method. Additionally, a fifth-order Magnus expansion (ME) is used to directly resolve the ODE (Eq.~\eqref{eq:wave_eq}), providing another validation of the analytical AE solutions.

\begin{figure}[!ht]
	\centering
	\subfigure
  		{
\includegraphics[width=6cm]{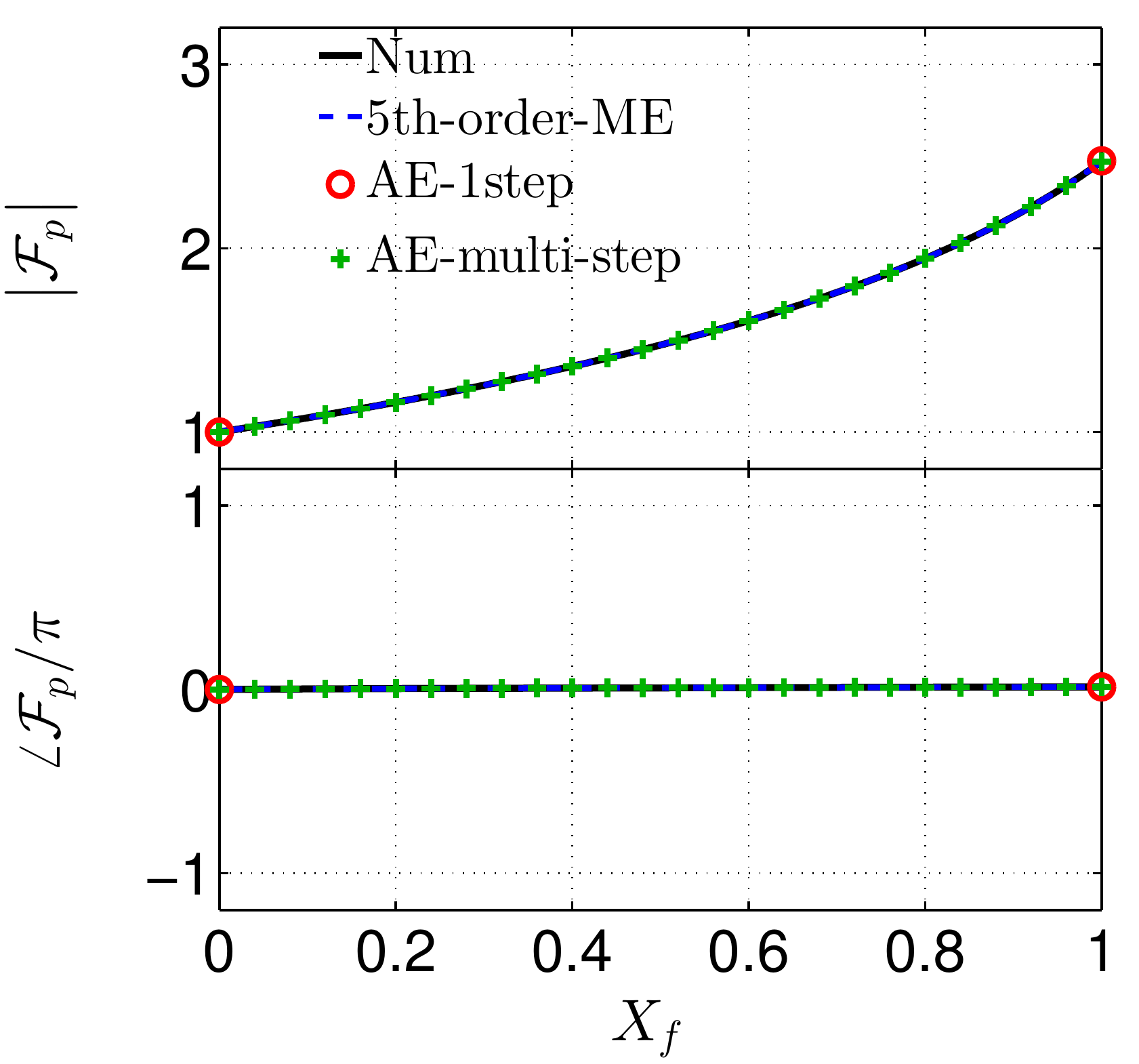}
  		}
  		\put (-190,150) {\normalsize$\displaystyle(a)$}
  		\vspace*{-0pt}
  		\hspace*{20pt}
  		\subfigure
  		  		{
\includegraphics[width=6cm]{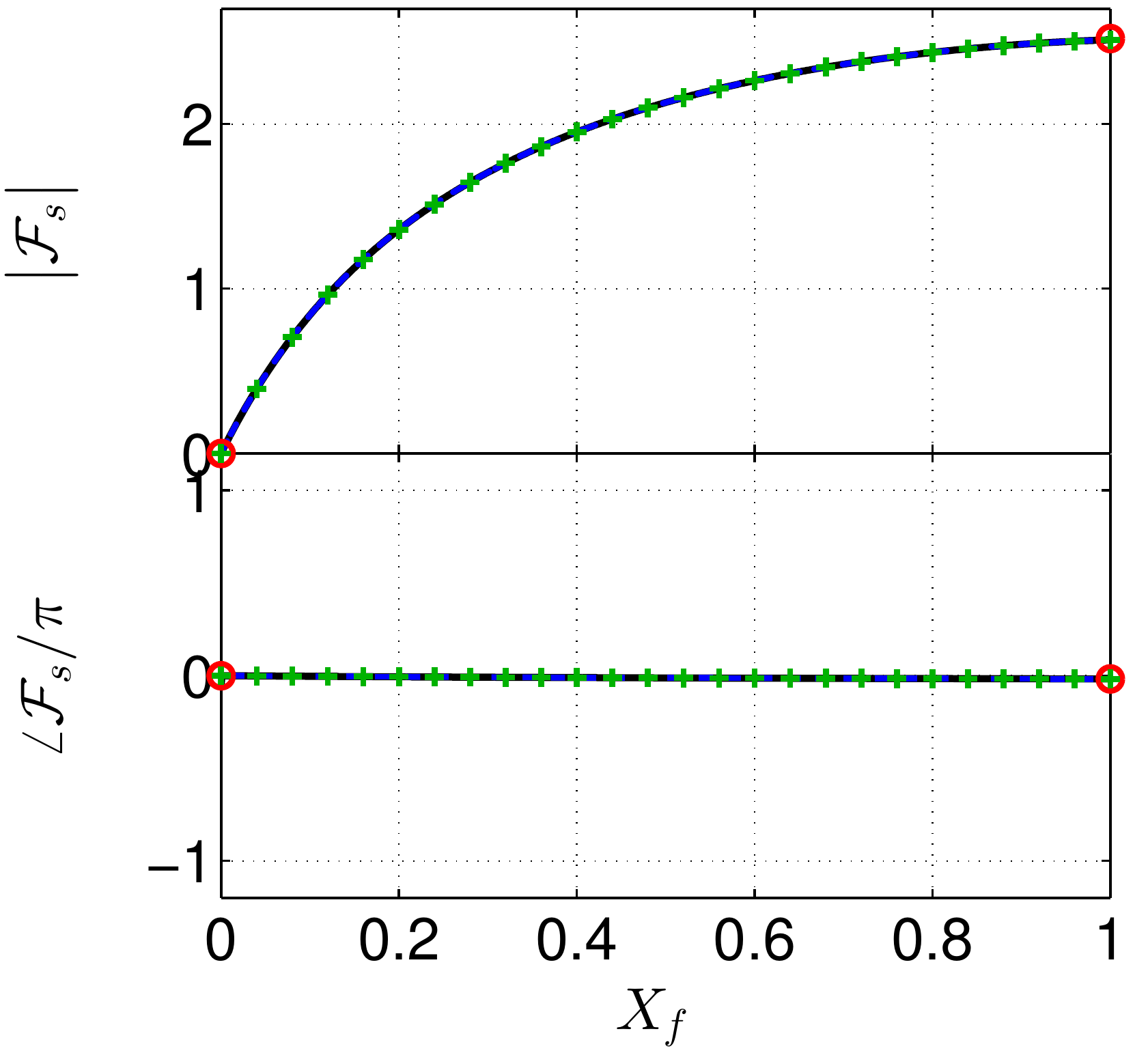}
  		}
  		\put (-190,150) {\normalsize$\displaystyle(b)$}
  		\vspace*{-0pt}
 	 \caption{Transfer functions (a) $Re(\mathcal{F}_p)$ and (b) $Re(\mathcal{F}_{s})$ as a function of axial coordinate, $X_f$, within the heat transfer region with $\Omega/2\pi=0.10$ and $M_u = 0.20$: numerical LEE simulation ($\mathbf{-}$), 5th-order ME method (\textcolor{blue}{$\mathbf{- -}$}), one-step AE method (\textcolor{red}{$\mathbf{\circ}$}) and multi-step AE method (\textcolor[rgb]{0,0.7,0}{$\mathbf{+}$}) are presented.}
	 \label{Fig:TF_f001_M020}
	 \vspace*{00pt}
\end{figure}   
\begin{figure}[!ht]
	\centering
	\subfigure
  		{
\includegraphics[width=6cm]{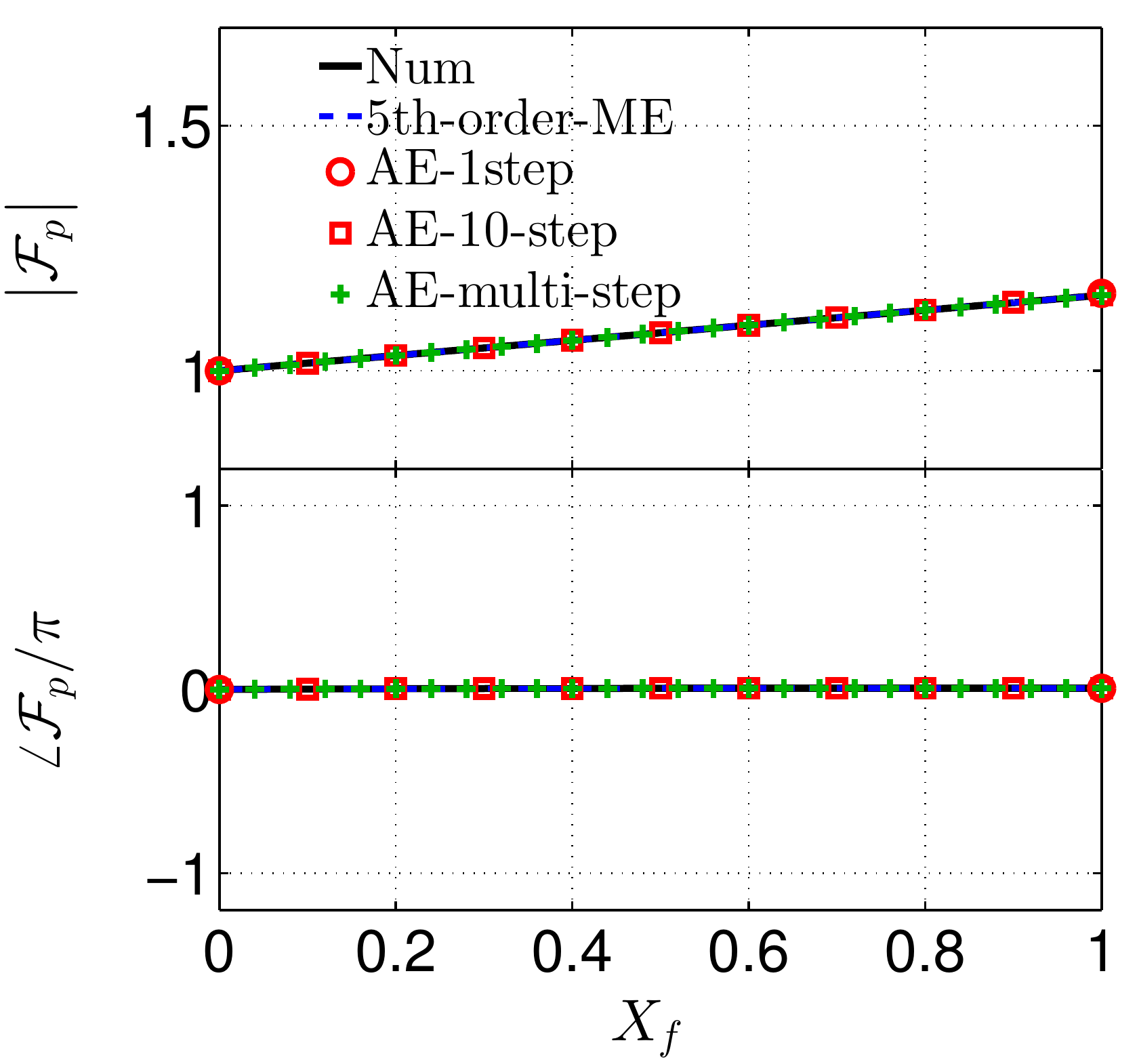}
  		}
  		\put (-190,150) {\normalsize$\displaystyle(a)$}
  		\vspace*{-0pt}
  		\hspace*{20pt}
  		\subfigure
  		  		{
\includegraphics[width=6cm]{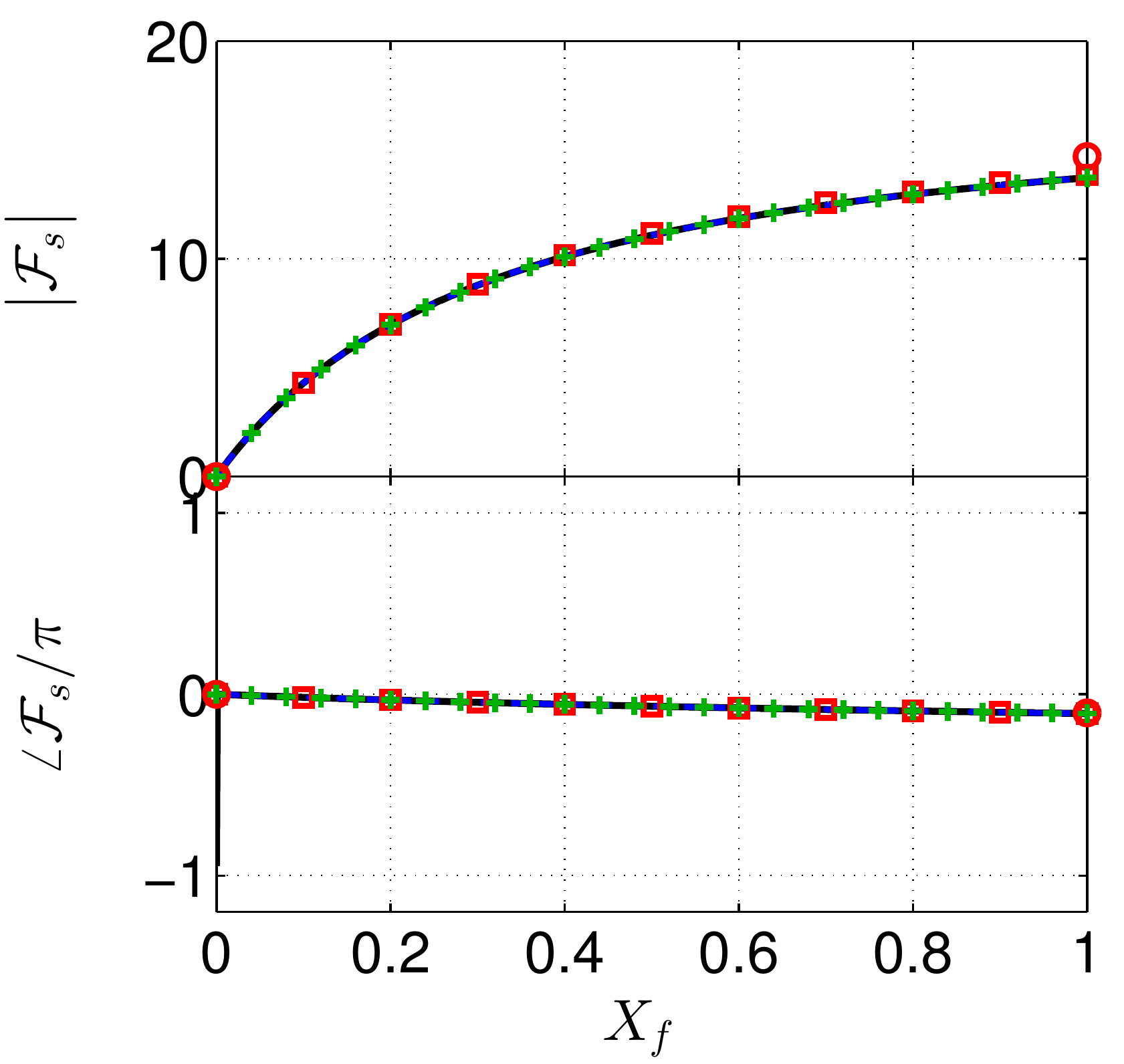}
  		}
  		\put (-190,150) {\normalsize$\displaystyle(b)$}
  		\vspace*{-0pt}
 	 \caption{Transfer functions (a) $Re(\mathcal{F}_p)$ and (b) $Re(\mathcal{F}_{s})$ as a function of axial coordinate, $X_f$, within the heat transfer region with $\Omega/2\pi=0.10$ and $M_u = 0.05$: numerical LEE simulation ($\mathbf{-}$), 5th-order ME method (\textcolor{blue}{$\mathbf{- -}$}), one-step AE method (\textcolor{red}{$\mathbf{\circ}$}), 10-step AE method (\textcolor{red}{$\mathbf{\square}$}) and multi-step AE method (\textcolor[rgb]{0,0.7,0}{$\mathbf{+}$}) are presented.}
	 \label{Fig:TF_f001_M005}
	 \vspace*{00pt}
\end{figure}
\begin{figure}[!ht]
	\centering
	\subfigure
  		{
\includegraphics[width=6cm]{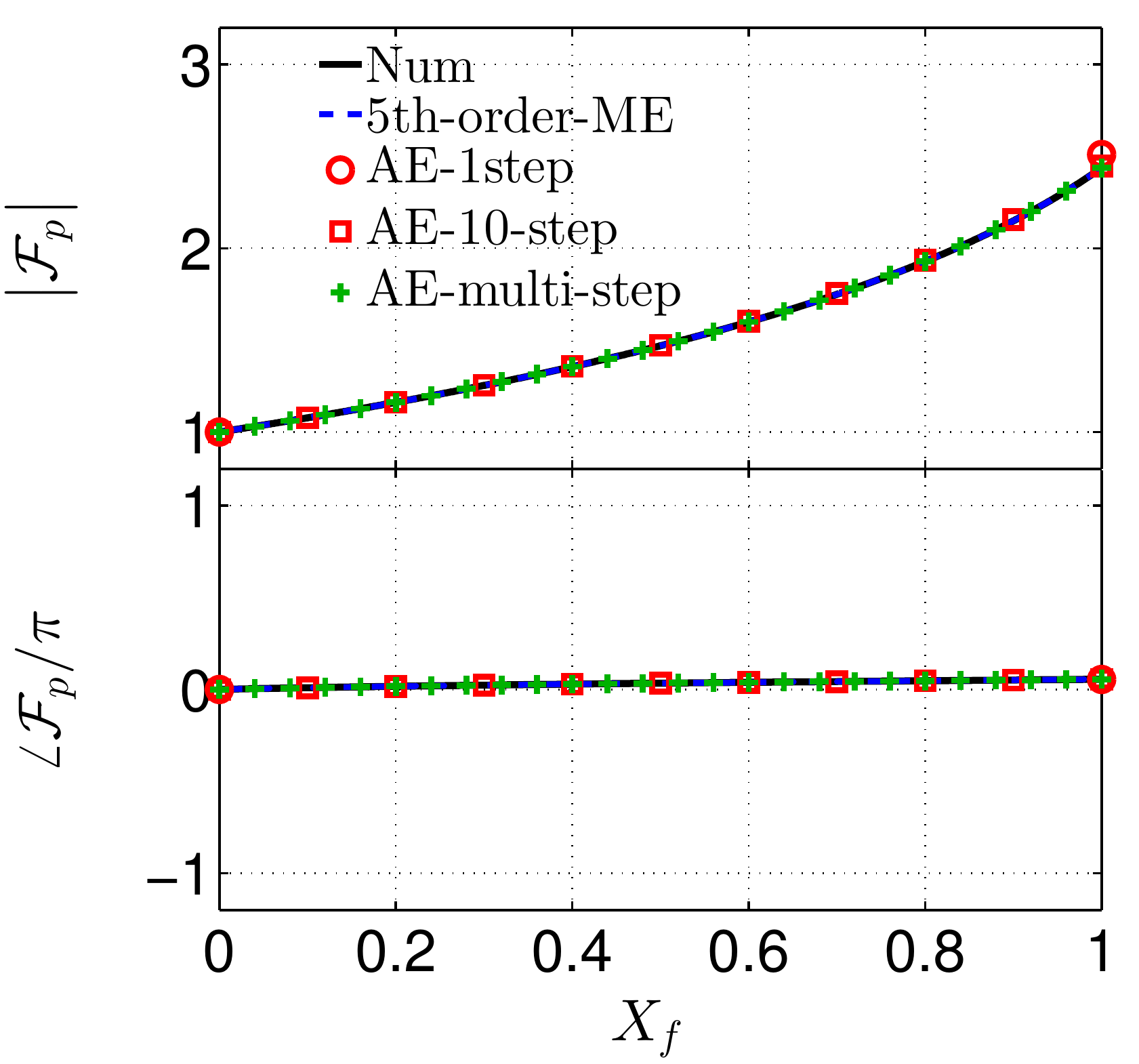}
  		}
  		\put (-190,150) {\normalsize$\displaystyle(a)$}
  		\vspace*{-0pt}
  		\hspace*{20pt}
  		\subfigure
  		  		{
\includegraphics[width=6cm]{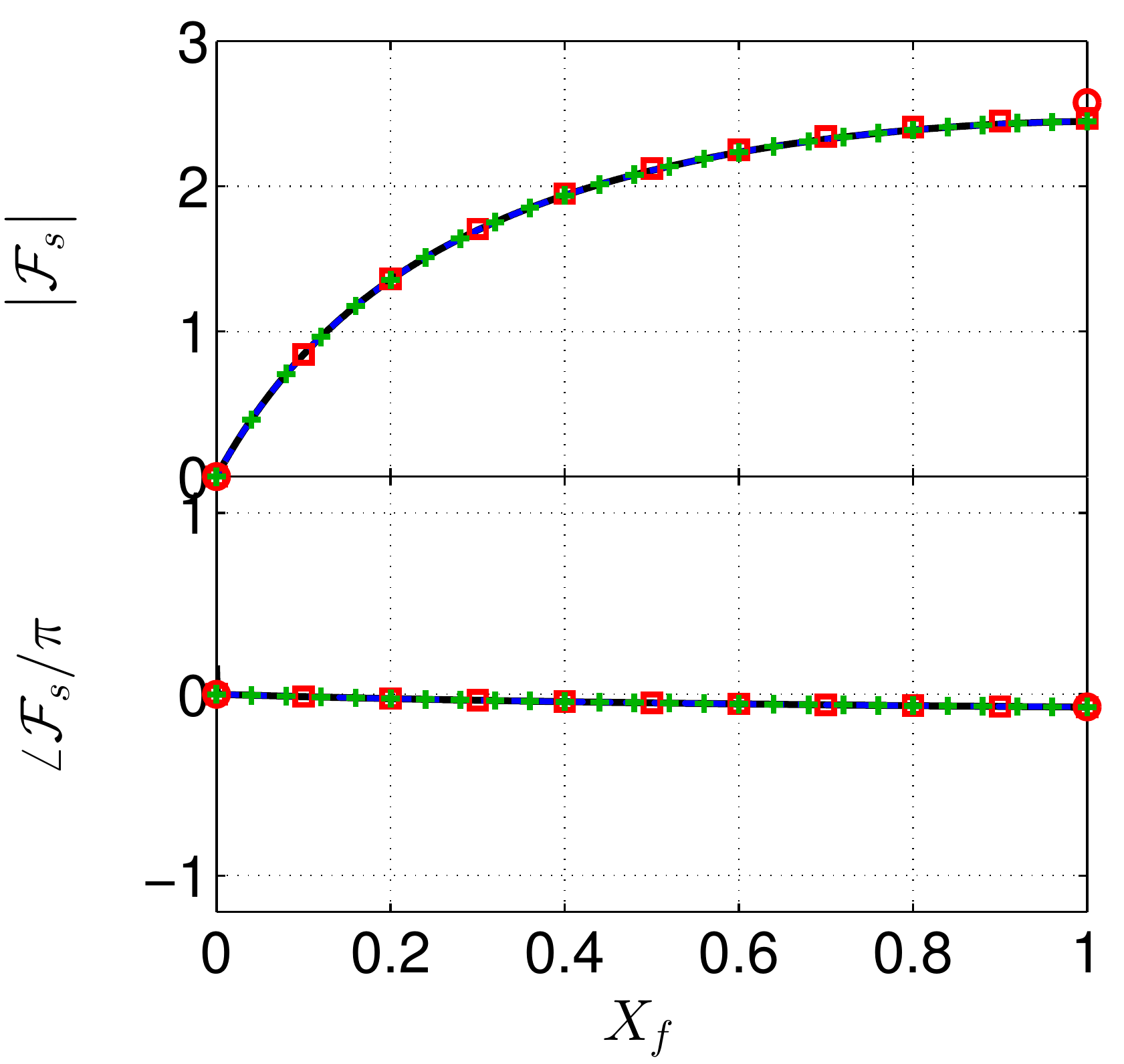}
  		}
  		\put (-190,150) {\normalsize$\displaystyle(b)$}
  		\vspace*{-0pt}
 	 \caption{Transfer functions (a) $Re(\mathcal{F}_p)$ and (b) $Re(\mathcal{F}_{s})$ as a function of axial coordinate, $X_f$, within the heat transfer region with $\Omega/2\pi=0.40$ and $M_u = 0.20$: numerical LEE simulation ($\mathbf{-}$), 5th-order ME method (\textcolor{blue}{$\mathbf{- -}$}), one-step AE method (\textcolor{red}{$\mathbf{\circ}$}), 10-step AE method (\textcolor{red}{$\mathbf{\square}$}) and multi-step AE method (\textcolor[rgb]{0,0.7,0}{$\mathbf{+}$}) are presented.}
	 \label{Fig:TF_f004_M020}
	 \vspace*{00pt}
\end{figure}
\begin{figure}[!ht]
	\centering
	\subfigure
  		{
\includegraphics[width=6cm]{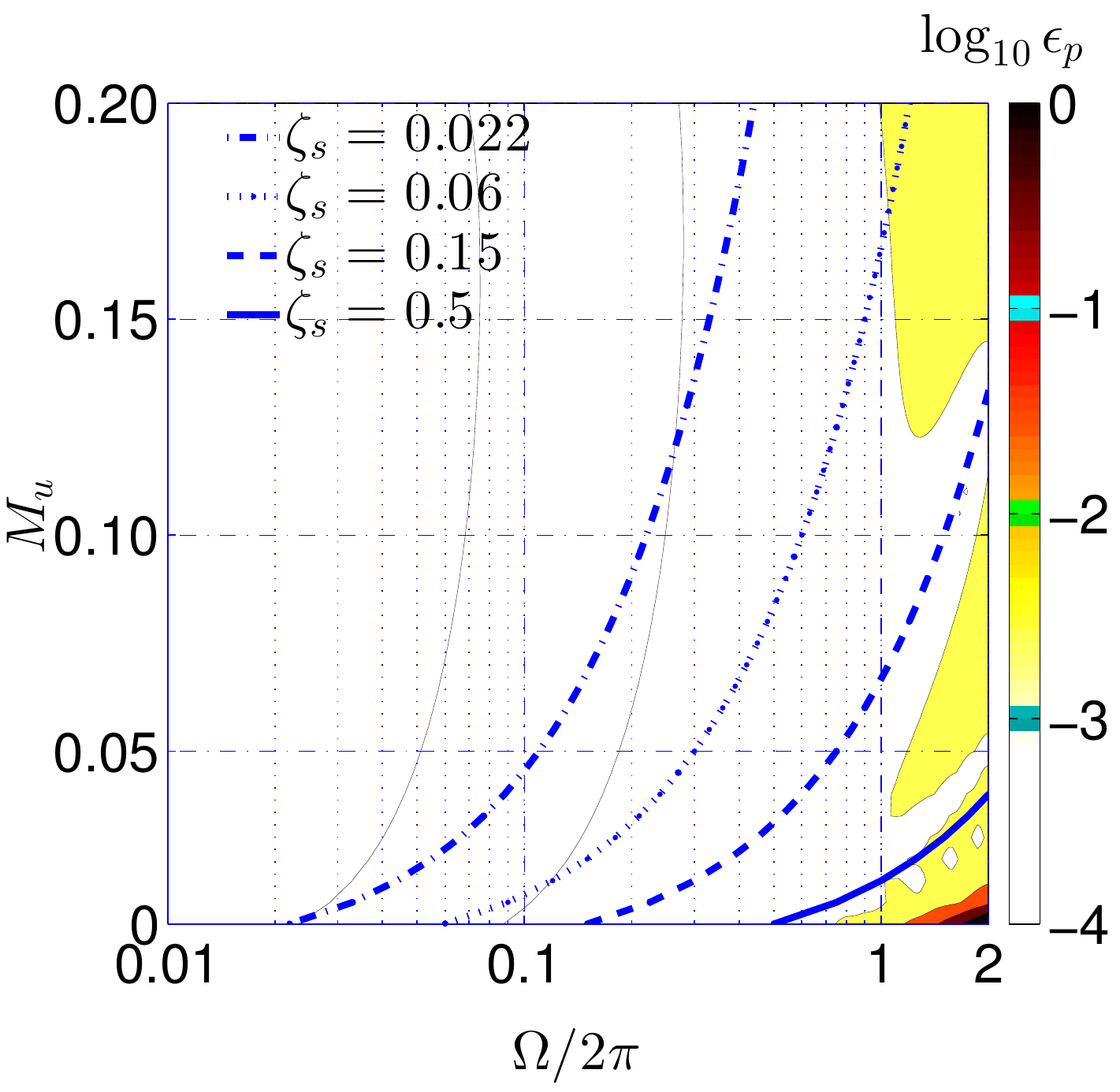}
  		}
  		\put (-190,150) {\normalsize$\displaystyle(a)$}
  		\vspace*{-0pt}
  		\hspace*{20pt}
  		\subfigure
  		  		{
\includegraphics[width=6cm]{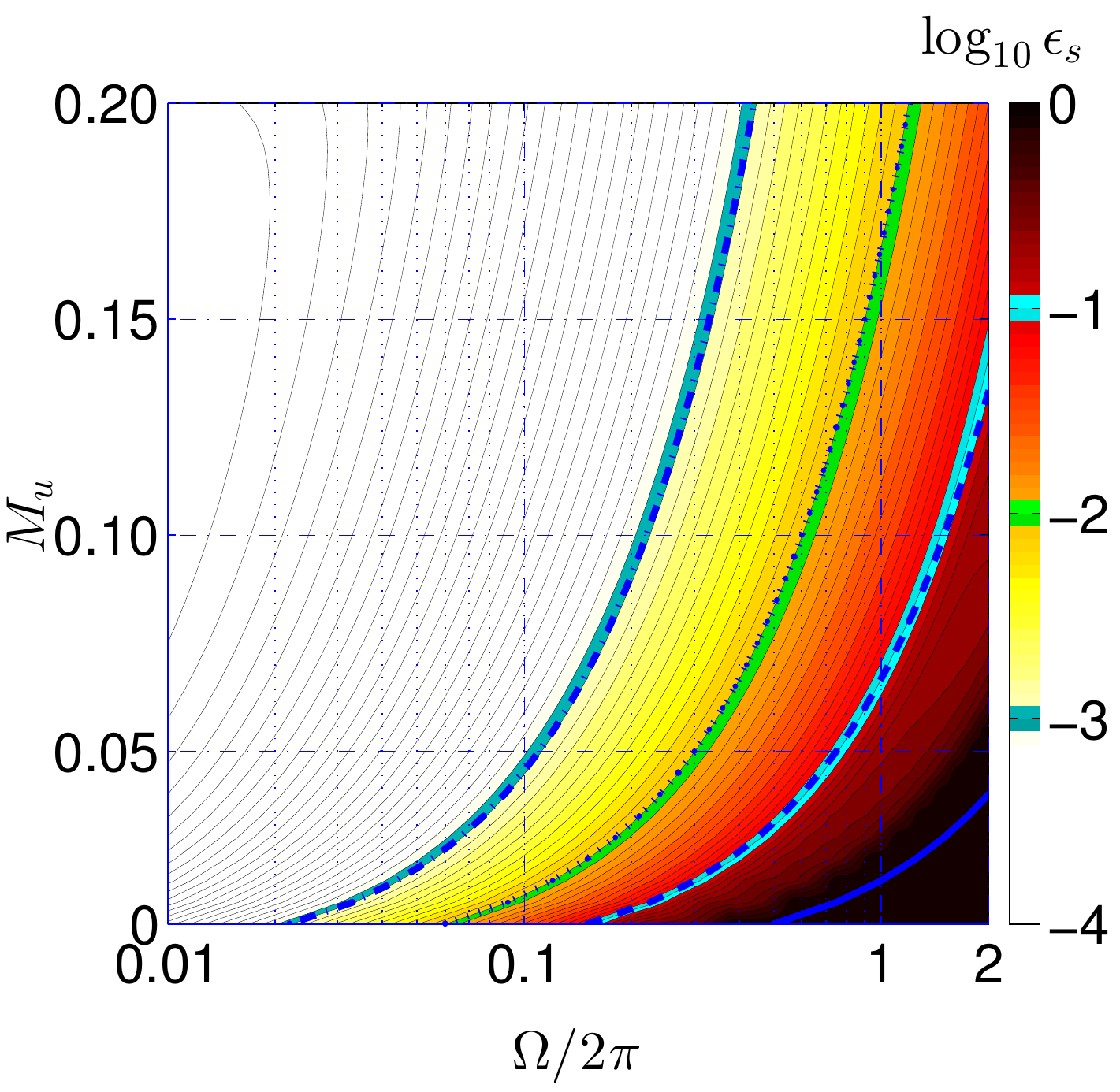}
  		}
  		\put (-190,150) {\normalsize$\displaystyle(b)$}
  		\vspace*{-0pt}
  		\caption{Contour maps for the error coefficients (a) pressure $\epsilon_p$ and (b) entropy $\epsilon_s$ using the multi-step AE method with $N_s=100$. Low-frequency conditions, $\Omega=\zeta_s \cdot \Omega_{LF}$, are plotted for different $\zeta_s$.  $\bar{T}_u=300$K, $\bar{T}_d=1200$K.}
	 \label{Fig:Contour_AE_100steps}
	 \vspace*{00pt}
\end{figure} 

A comparison of the transfer functions for the case of $\Omega/2 \pi = 0.10$ and $M_u=0.20$ is shown in Fig.~\ref{Fig:TF_f001_M020}. The one-step AE solution is able to predict the transfer functions at the exit of the heating region in excellent agreement with the numerical LEE simulations and the ME method. The multi-step AE method is also seen to offer excellent agreement.

Figure \ref{Fig:TF_f001_M005} compares the transfer functions for the same frequency, $\Omega/2 \pi = 0.10$, but a smaller Mach number value of $M_u=0.05$. The one-step prediction of the pressure transfer function still agrees well with the numerical LEE results and ME method. However, the entropy transfer function at the exit of the heating region exhibits an error, this depending on the ratio $\Omega/M_u=4\pi$. According to $\zeta_s$ and $\zeta_a$ defined in Sec.\ref{subsec:AE_steps}, $\zeta_a$ is independent of the flow Mach number, while $\zeta_s$ is proportional to $\Omega/M$. The multi-step AE method with 10 uniform segments is able to correctly calculate the transfer functions at the region exit.

The case of a larger frequency, $\Omega/2 \pi = 0.40$ and $M_u=0.20$ ($\Omega/M_u=4\pi$) is presented in Fig.~\ref{Fig:TF_f004_M020}. The increase in frequency results in the one-step AE method being less accurate for both the acoustic and entropy transfer functions. The 10-step AE method corrects the errors, showing good match to predictions from the multi-step AE method, the numerical LEE results and the ME method.

The multi-step AE solutions using $N_s=100$ steps are further evaluated by considering the pressure and entropy error coefficients as functions of the normalised frequency, $\Omega$, and upstream flow Mach number, $M_u$, in Fig.~\ref{Fig:Contour_AE_100steps}. Four different values of $\zeta_s$ are selected for the low-frequency condition to calculate four limit lines of $\Omega=\zeta_s \cdot \Omega_{\mathrm{LF}}$. It is seen that the smaller the value of $\zeta_s$ (i.e. the further the imposed frequency, $\Omega$, is from the low-frequency condition $\Omega_{\mathrm{LF}}$), the higher the accuracy. The pressure transfer functions are seen to be accurately predicted over almost the whole region except for the lower right corner of higher frequencies and smaller Mach numbers. The entropy transfer function exhibits poor accuracy around the limit line $\zeta_s=0.5$, indicating that accurate prediction of the entropy wave is what limits the validity of the AE method for higher frequencies and smaller Mach numbers. It is therefore ensured that the multi-step AE solutions are conducted with sufficient segments in the following analyses.

\subsection{Eigenvalue system of the duct with the finite-length passive flame}
\label{sec:EigenFre_with_flame_thickness}
\begin{table}[!ht]
\caption{The first three thermoacoustic modes of the eigenvalue system with different passive flame thicknesses.}\vskip1ex
\centering
\label{para_table}
\begin{tabular}{ccccc}
\toprule
 	  &     &\multicolumn{3}{c}{$M_u=0.1,~b/L=0.25$} \\
\midrule
$\delta_f/L$ [-]	  & LEEs/AE      &$f_1+\mathrm{i}{Gr}_1$ [$s^{-1}$]  &$f_2+\mathrm{i}{Gr}_2$ [$s^{-1}$]     &$f_3+\mathrm{i}{Gr}_3$ [$s^{-1}$]\\
\midrule
\multirow{2}*{compact}   &\multirow{2}*{--} 	&\multirow{2}*{$104.6-128.1\mathrm{i}$}  &\multirow{2}*{$251.4+50.8\mathrm{i}$} &\multirow{2}*{$376.7+31.0\mathrm{i}$}\\
                         &    &  &  &\\
\midrule
\multirow{2}*{0.01}   &LEEs 	&\multirow{2}*{$103.9-129.2\mathrm{i}$ }  &\multirow{2}*{$250.9+45.9\mathrm{i}$} 
&\multirow{2}*{$374.8+35.0\mathrm{i}$ } \\
                         &AE        &  &  & \\
                         \midrule
\multirow{2}*{0.05}   &LEEs 	&$101.4-134.0\mathrm{i}$ &$248.7+23.8\mathrm{i}$  &$366.7+43.6\mathrm{i}$ \\
                      &AE    &$101.4-133.9\mathrm{i}$  &$248.7+23.9\mathrm{i}$  &$366.7+43.9\mathrm{i}$   \\
\midrule
\multirow{2}*{0.1}   &LEEs 	&$98.5-140.4\mathrm{i}$ &$246.1 - 10.2\mathrm{i}$
&$355.4+36.1\mathrm{i}$ \\
                     &AE    &$98.5-140.3\mathrm{i}$  &$246.0 - 9.8\mathrm{i}$
 &$355.4+37.3\mathrm{i}$ \\
\bottomrule
\end{tabular}\\
\vspace{0.5em}
\footnotesize{$*$ The negative value of $Gr$ indicates the spatially damping wave of the corresponding resonant frequency $f$. }\\
\footnotesize{$*$ $N_s=300$ is chosen to ensure that the biggest eigenfrequencies $f_3$ are accurately captured in all cases.}\\
\end{table}

In this subsection, the proposed multi-step AE solutions are applied to a duct containing a distributed passive flame, and are used to predict the thermoacoustic modes. The duct inlet is acoustically closed, $\hat{\boldsymbol{U}}(X=0)=0$, and the outlet is choked, 
\begin{equation}
\label{eq:choked_BC}
- \frac{\hat{\boldsymbol{P}}}{\bar{\boldsymbol{P}}} +\frac{\hat{\boldsymbol{\rho}}}{\bar{\boldsymbol{\rho}}} +  \frac{2 \hat{\boldsymbol{U}}}{ \bar{\boldsymbol{U}}}\, \Bigg|_{X=1}=0.
\end{equation}
Due to their uniform temperature profiles, the regions both upstream and downstream of the distributed passive flame have direct analytical solutions for the acoustic and entropy waves, as presented in Eqs.~\eqref{eq:Sol_p_mean_T}-\eqref{eq:Sol_u_mean_T}. The distributed flame region is processed by the AE solutions to connect the upstream and downstream regions.

The eigenvalue system is consequently built based upon the acoustic transfer matrix of the AE method and both boundary conditions of duct ends, expressed as: 
\begin{equation}
\label{eq:AE_Eigen}
\mathbb{M}^{\mathrm{AE}} \cdot C^{\mathrm{AE}} =
\left[
\begin{array}{c|c}
\begin{matrix}
   \vspace{0.2cm}\\
 \left[\boldsymbol{\mathrm{T}}\right]_{u}^{d}
\mathcal{M}^{w2p}_u \vspace{0.2cm}\\
 \\
 \end{matrix} 
 &
 \begin{matrix}
 -\mathcal{M}^{w2p}_d 
 \end{matrix} \\ \hline
 \begin{matrix}
   &  \vspace{-0.2cm}\\
  1/\bar{\boldsymbol{\rho}}_u \bar{\boldsymbol{c}}_u
&-1/\bar{\boldsymbol{\rho}}_u \bar{\boldsymbol{c}}_u \vspace{0.2cm}\\
 0 & 0\\
 \end{matrix}
 &\begin{matrix}
   & & \vspace{-0.2cm}\\
 0 &0 &0 \vspace{0.2cm}\\
  R_1 &R_2 &R_3 \\
 \end{matrix} 
\end{array}
\right]_{5\times 5}
 \begin{bmatrix}
A_u^+ \vspace{0.2cm}\\
A_u^- \vspace{0.2cm}\\
A_d^+ \vspace{0.2cm}\\
A_d^- \vspace{0.2cm}\\
 E_d\\
\end{bmatrix}
 = \mathbf{0}\,.
\end{equation}
The derivation of this eigenvalue system is presented in detail in \ref{sec:eigenvalue}.

Replacing $[\mathrm{T}]_{u}^{d}$ in Eq.~\eqref{eq:AE_Eigen} by the compact acoustic transfer matrix $[\mathrm{T}_c]_{u}^{d}$ leads to the eigenvalue system of the compact passive flame:
\begin{equation}
\begin{matrix}
\label{eq:compact_Eigen}
\mathbb{M}^{\mathrm{c}} =
\left[
\begin{array}{c|c}
\begin{matrix}
   \vspace{0.2cm}\\
 \left[\boldsymbol{\mathrm{T}}_c\right]_{u}^{d}
\mathcal{M}^{w2p}_u \vspace{0.2cm}\\
 \\
 \end{matrix} 
 &
 \begin{matrix}
 -\mathcal{M}^{w2p}_d 
 \end{matrix} \\ \hline
 \begin{matrix}
   &  \vspace{-0.2cm}\\
  1/\bar{\boldsymbol{\rho}}_u \bar{\boldsymbol{c}}_u
&-1/\bar{\boldsymbol{\rho}}_u \bar{\boldsymbol{c}}_u \vspace{0.2cm}\\
 0 & 0\\
 \end{matrix}
 &\begin{matrix}
   & & \vspace{-0.2cm}\\
 0 &0 &0 \vspace{0.2cm}\\
    R_1 &R_2  &R_3 \\
 \end{matrix} 
\end{array}
\right]_{5\times 5}\,.
\end{matrix}
\end{equation}

To validate the performance of the AE method, the LEEs are applied to the thermoacoustic system by dividing the distributed region into the same number of subregions as the multi-step AE method. A large number of discrete wave amplitude coefficients of $C^{\mathrm{LEEs}}$ consequently result in a large square matrix $\mathbb{M}^{\mathrm{LEEs}}$ for the eigenvalue system for the LEEs method, as expressed in Eqs.~\eqref{eq:LEE_C}-\eqref{eq:LEE_Eigen}. 
\begin{equation}
\label{eq:LEE_C}
C^{\mathrm{LEEs}} = \Big[
A_u^+, A_u^-,  \underbrace{\cdots, A_{f,i}^+, A_{f,i}^-, E_{f,i},\cdots,}_{i=1,\cdots,N_s}  A_d^+,  A_d^-,  E_d \Big]^{\mathrm{T}}
\end{equation}

\begin{equation}
\label{eq:LEE_Eigen}
\mathbb{M}^{\mathrm{LEEs}} \cdot C^{\mathrm{LEEs}}
 = 0
\end{equation}

The thermoacoustic modes of each eigenvalue system can then be calculated by setting the determinant of $\mathbb{M}$ to 0, written as:
\begin{equation}
\Big|\mathbb{M}(\omega)\Big|=0\,.
\end{equation}

\begin{figure}[!ht]
	\centering
	\subfigure
  		{
\includegraphics[width=5.1cm]{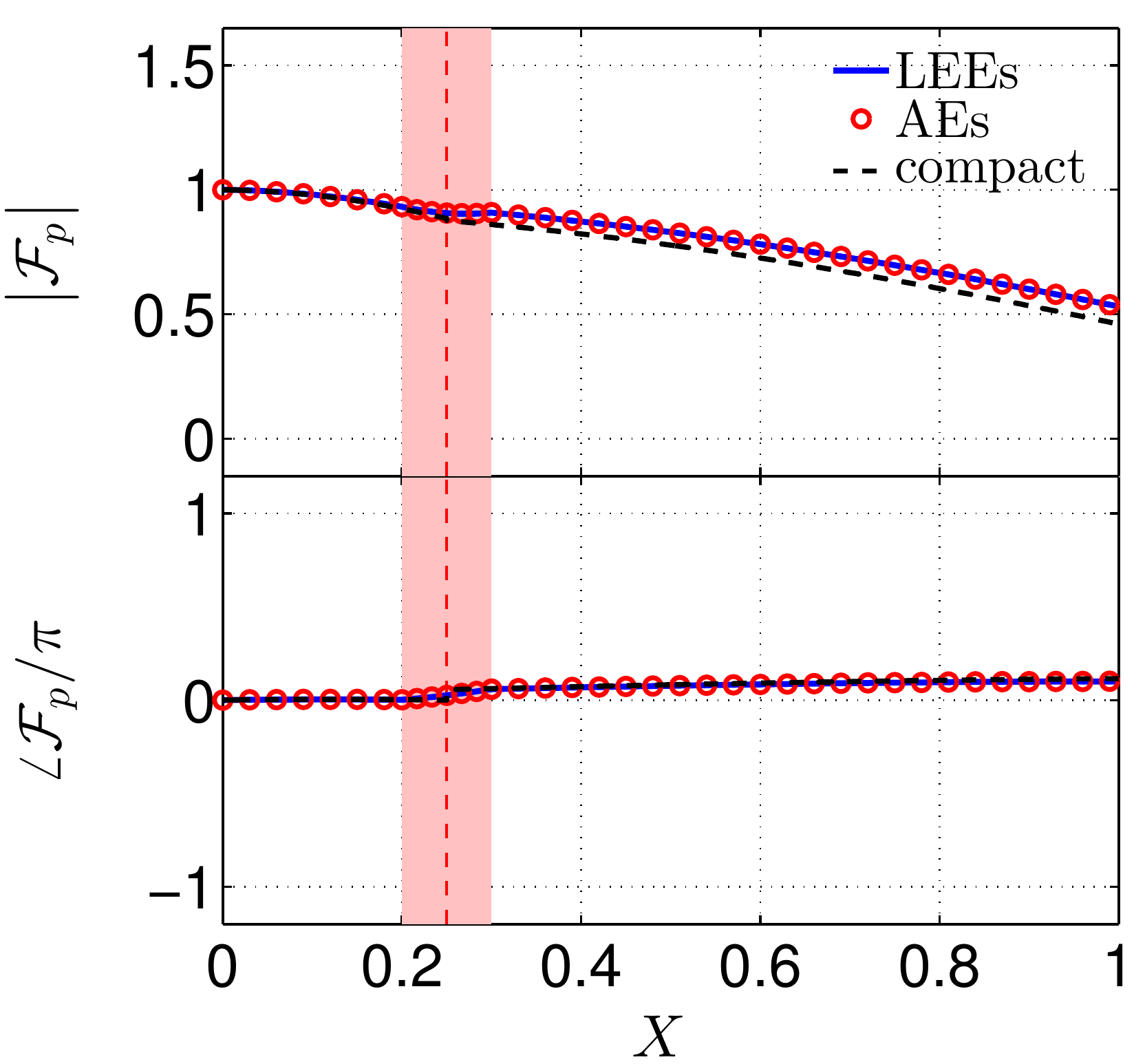}
  		}
  		\put (-152,125) {\normalsize$\displaystyle(a)$}
  		  \put (-80,136) {\normalsize First Mode}
  		\vspace*{-0pt}
  		\hspace*{2pt}
  		\subfigure
  		  		{
\includegraphics[width=5.1cm]{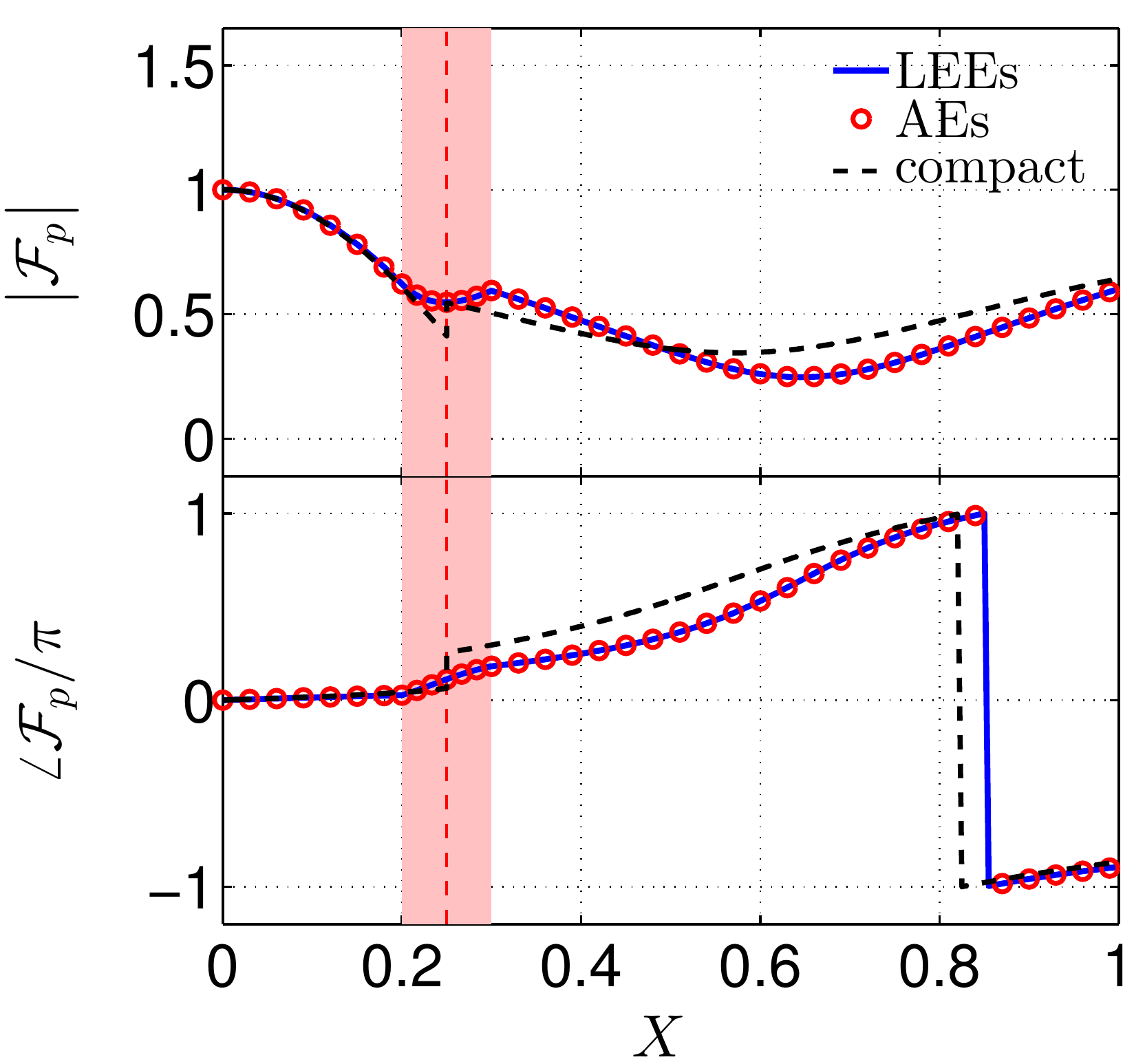}
  		}
  		\put (-152,125) {\normalsize$\displaystyle(b)$}
  		\put (-88,136) {\normalsize Second Mode}
  		\vspace*{0pt}
  		 \hspace*{2pt}
\subfigure
  		{
\includegraphics[width=5.1cm]{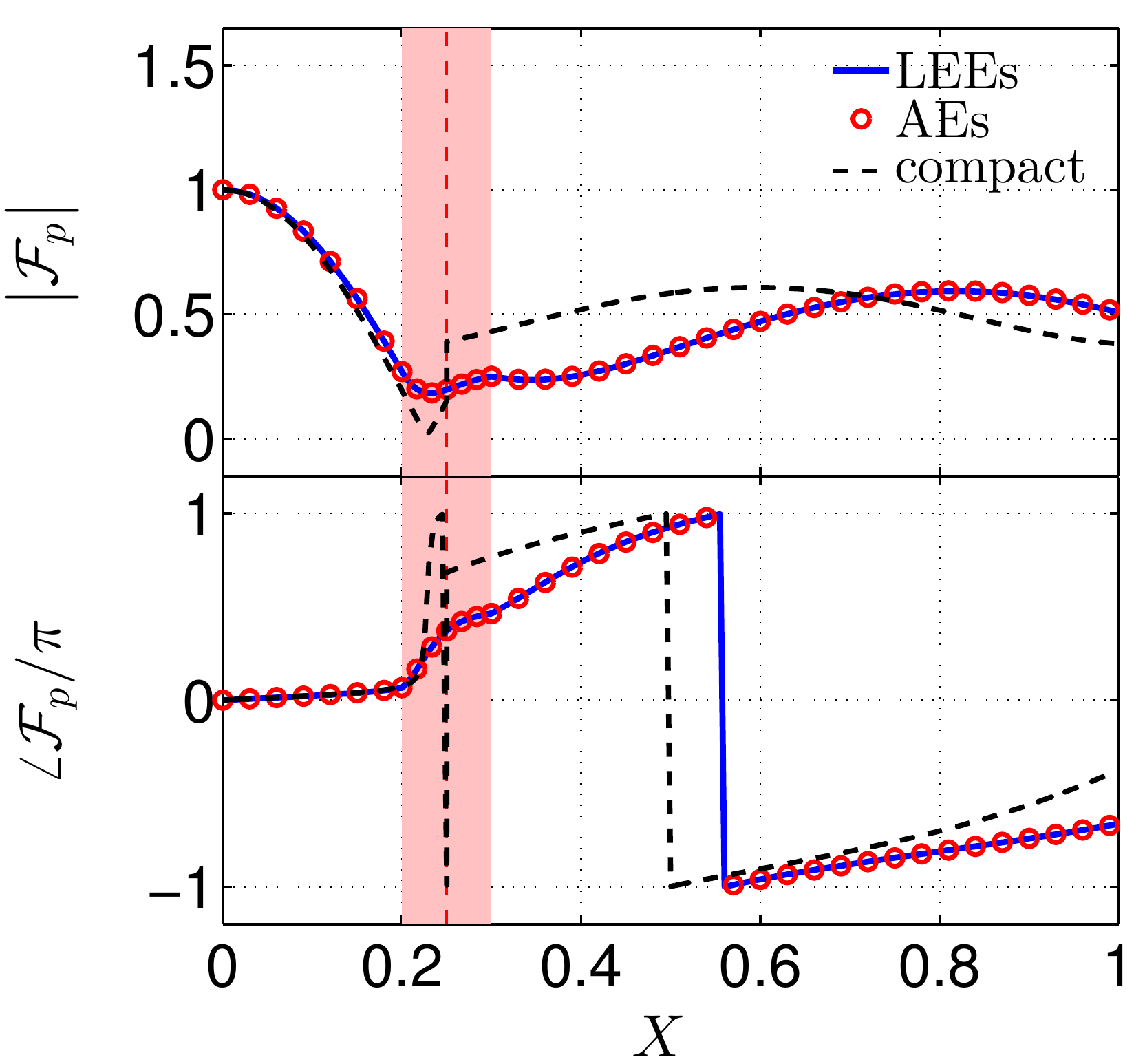}
  		}
  		\put (-152,125) {\normalsize$\displaystyle(c)$}
  		\put (-88,136) {\normalsize Third Mode}
  		\vspace*{-0pt}
  		\hspace*{-0pt}
  		\subfigure
  		  		{
\includegraphics[width=5.1cm]{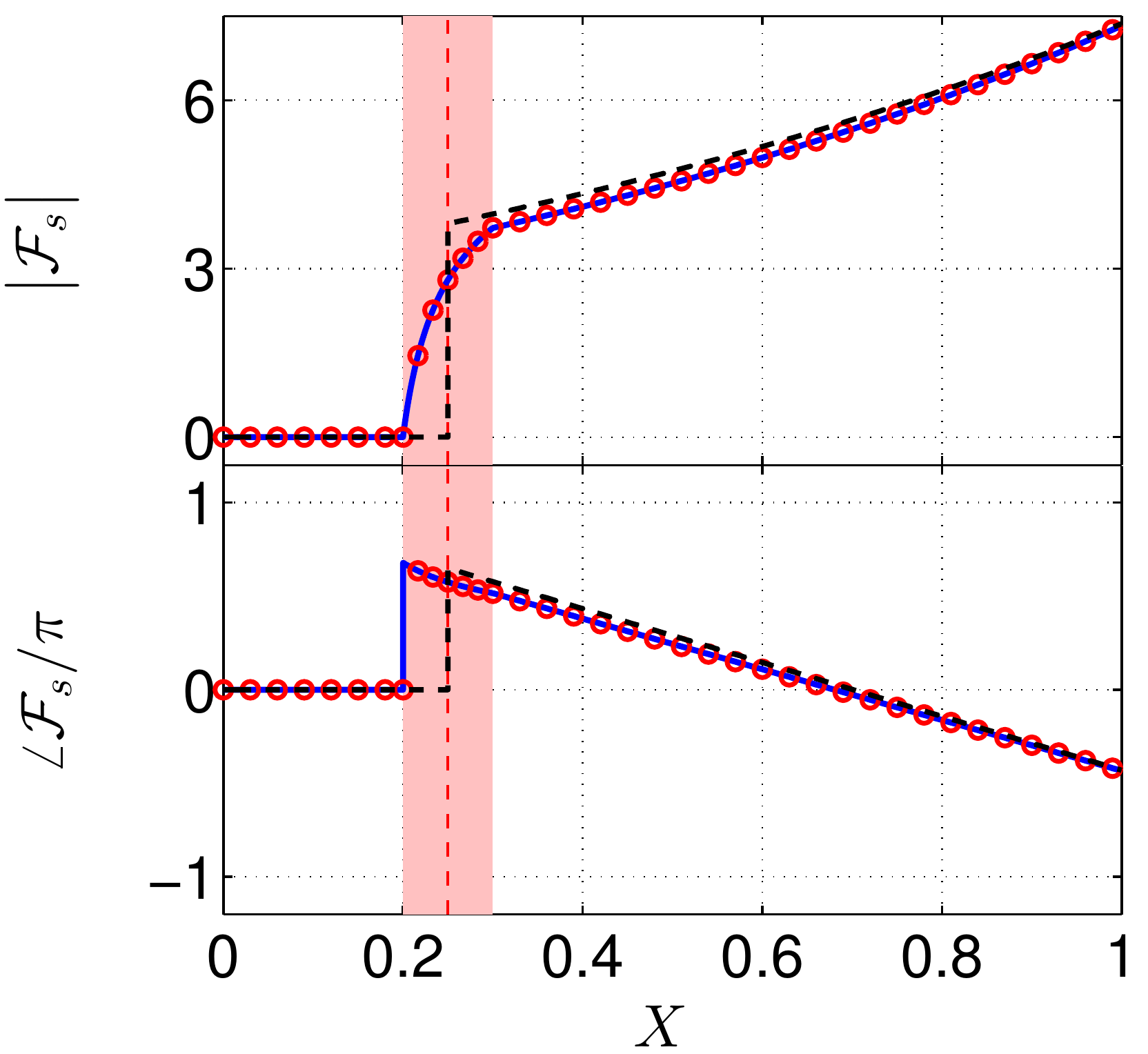}
  		}
  		\put (-152,125) {\normalsize$\displaystyle(d)$}
  		\vspace*{-0pt}
  	    \hspace*{2pt}
	\subfigure
  		{
\includegraphics[width=5.1cm]{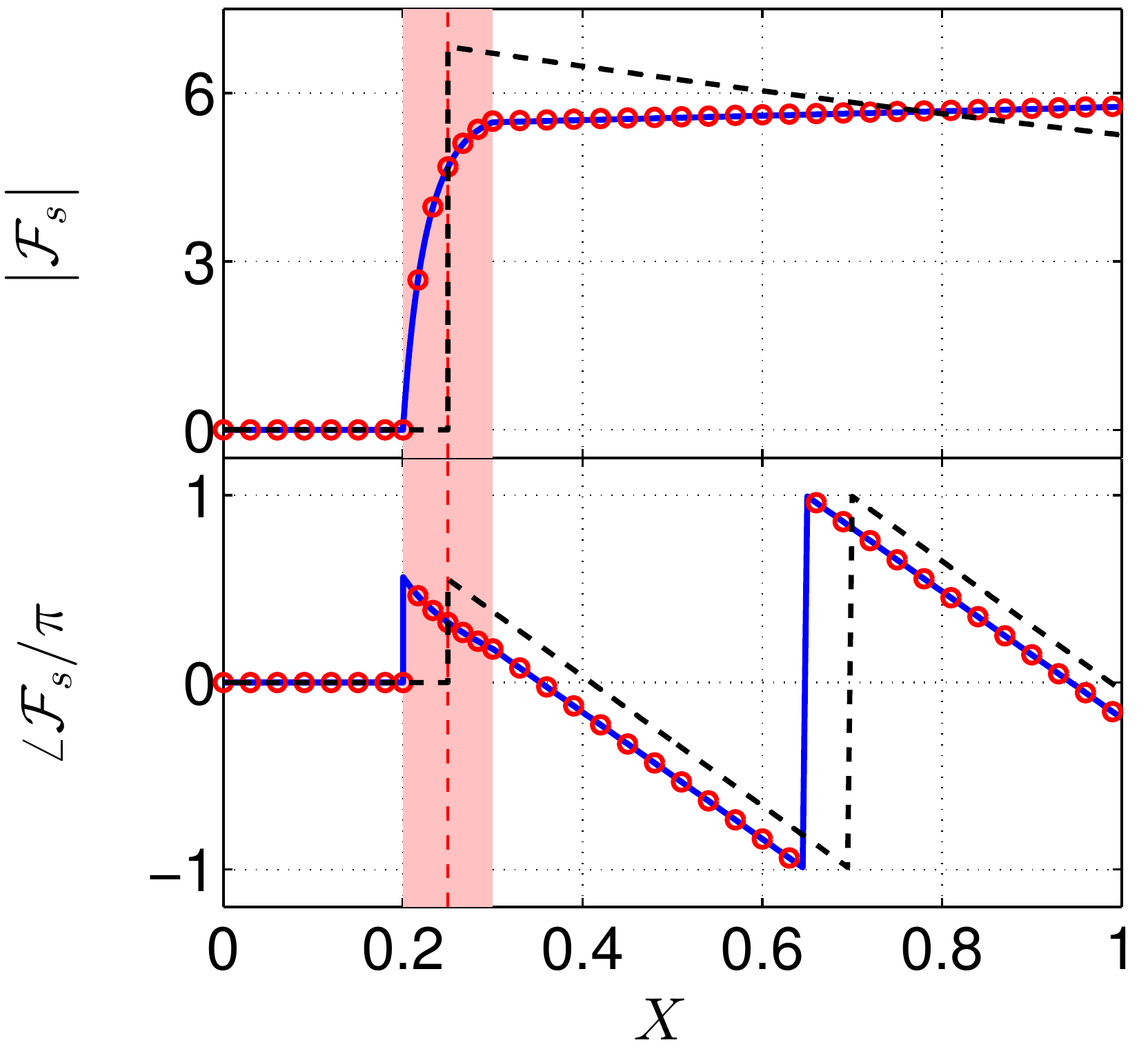}
  		}
  		\put (-152,125) {\normalsize$\displaystyle(e)$}
  		\vspace*{-0pt}
  		\hspace*{2pt}
  		\subfigure
  		  		{
\includegraphics[width=5.1cm]{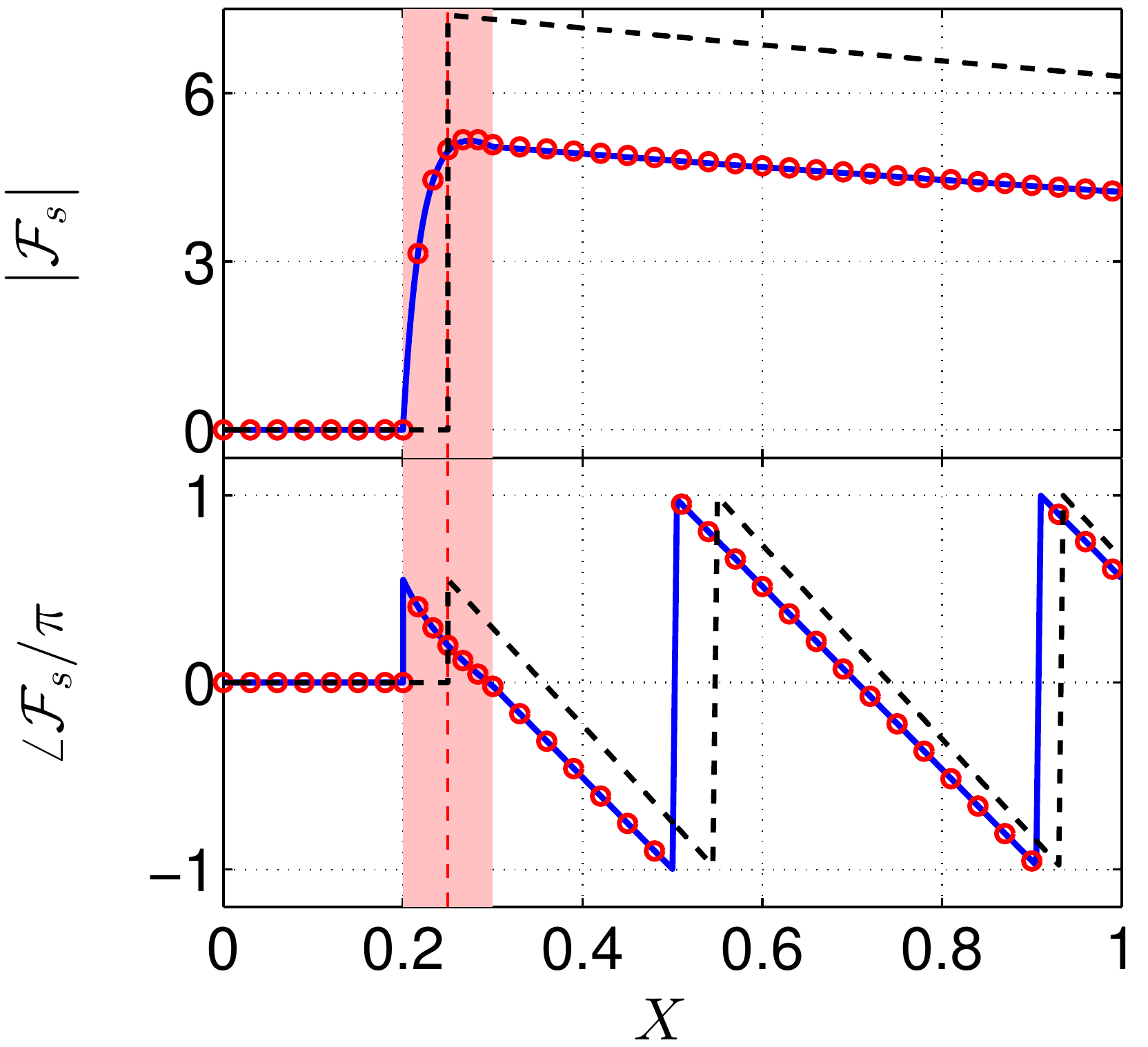}
  		}
  		\put (-152,125) {\normalsize$\displaystyle(f)$}
  		\vspace*{-0pt}  		
 	 \caption{Profiles of the (a-c) acoustic and (d-f) entropy transfer functions for the first three modes of the duct with the passive flame fixed at a quarter of the duct length and $M_u = 0.10$. The position of the compact flame or the centre of the distributed flame is represented by the red dotted line (\textcolor{red}{$\mathbf{--}$}), and the distributed flame region is represented by the red highlighted block. $\delta_f/L=0.1$, $\bar{T}_u=300$K, $\bar{T}_d=1200$K.}
	 \label{Fig:TF_Mode1_M010_b025}
	 \vspace*{00pt}
\end{figure}
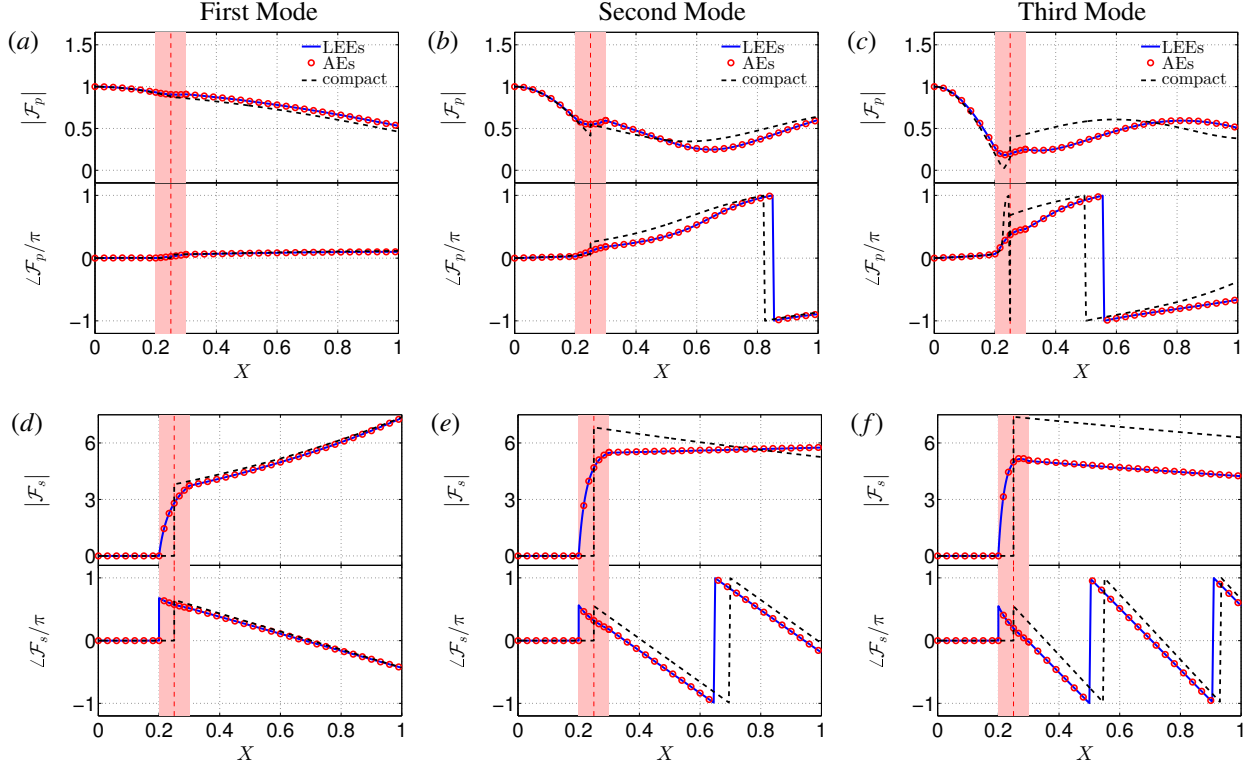  

The first three thermoacoustic modes are calculated with the centre of the passive flame fixed at a quarter of the duct length under the condition of $M_u=0.1$. Predictions by both the AE and LEE methods for compact and non-compact flames of $\delta_f/L=0, 0.01,0.05$ and $0.1$ are listed in Table \ref{para_table}. It should be noted that the AE and LEE methods have identical eigenvalue systems for the compact case. Comparisons show that AE solutions agree well with the LEE results for different steady flame thicknesses. Even with a thickness of $\delta_f/L=0.01$, the results of the distributed model exhibit differences from those of the compact one.
As the normalised flame thickness increases, the compact model yields increasingly inaccurate predictions for both eigenfrequencies and growth rates.

To further analyse the errors in the compact model, the acoustic and entropy transfer functions of the three modes are plotted with $\delta_f/L=0.1$ and $M_u=0.1$ in Fig.~\ref{Fig:TF_Mode1_M010_b025}. The predictions of the AE solutions are consistent with those from the LEEs and accurately reconstruct the corresponding mode structures of the acoustic and entropy waves. 
It can be seen that the acoustic and entropy waves of the non-compact cases exhibit different profiles from those of the compact case. The downstream entropy wave, generated by the interaction of incident acoustic waves upstream and the distributed mean temperature resulting from the passive flame, differs to that for the compact model. This entropy wave has a marked impact on the subsequent acoustic waves, especially when it meets a density- or entropy-dependent boundary condition at the duct end, such as the choked condition. 
Therefore, accounting for the spatial variation of the distributed steady heat source is essential for accurately predicting the modes of the thermoacoustic system.

\begin{figure}[!ht]
	\centering
\includegraphics[width=6cm]{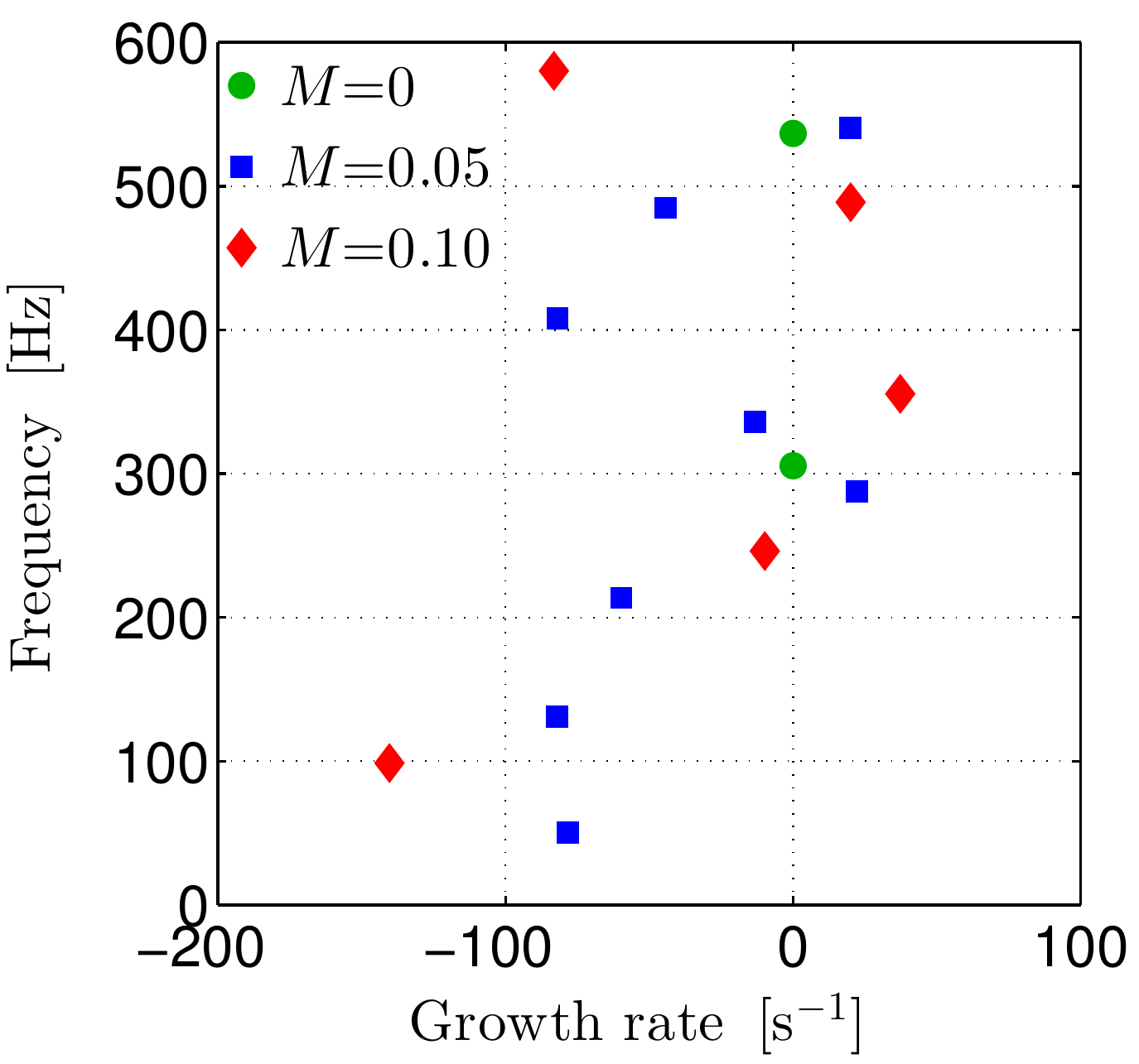}
  		\put (-190,150) {\normalsize$\displaystyle(a)$}
  		\vspace*{-0pt}
  		\hspace*{20pt}
  		\subfigure
  		  		{
\includegraphics[width=6cm]{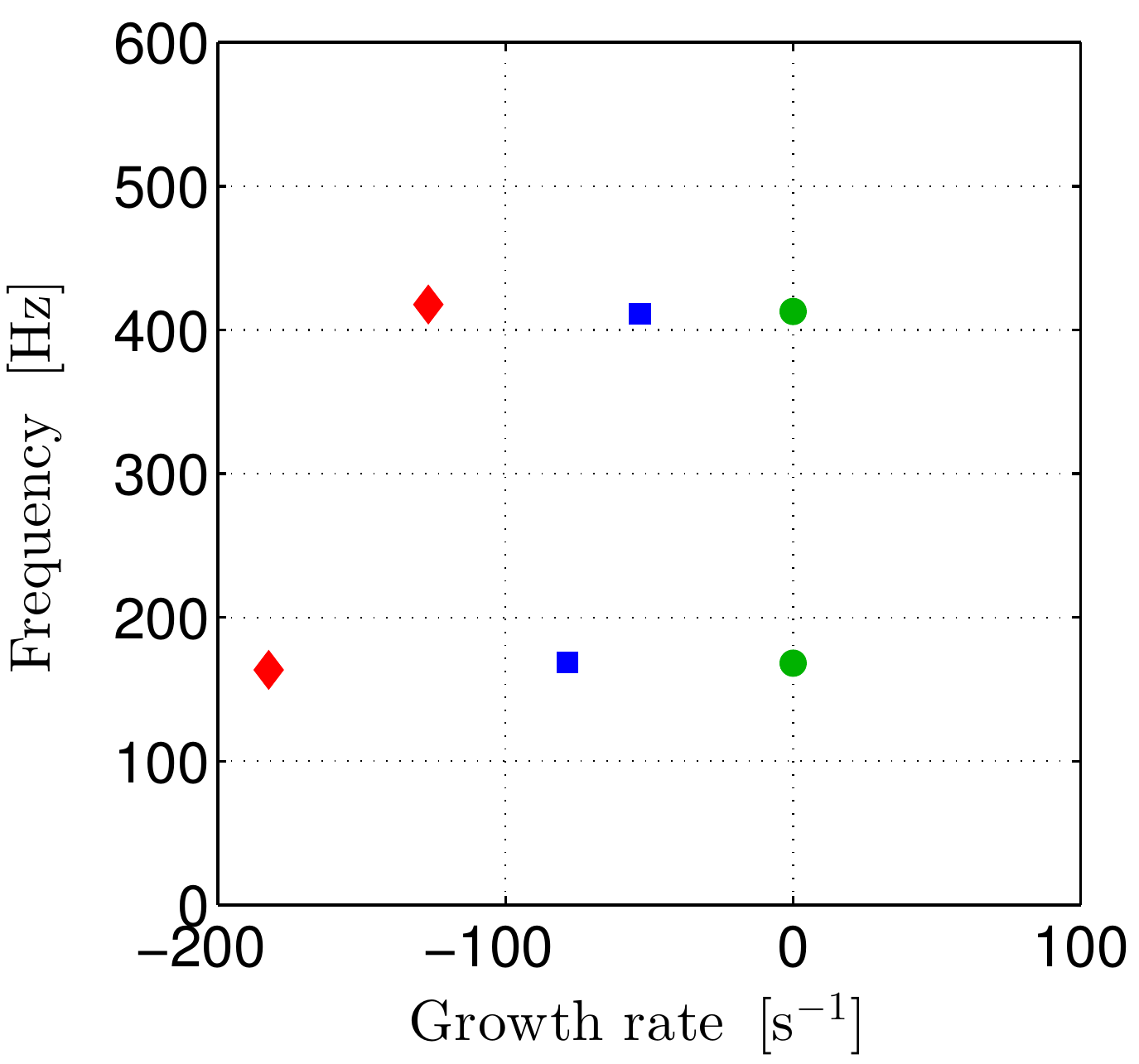}
  		}
  		\put (-190,150) {\normalsize$\displaystyle(b)$}
  		\vspace*{-0pt}
 	 \caption{Evolutions of the thermocoustic modes as a function of the incoming flow Mach numbers for a duct with (a) a choked boundary and (b) an open-end boundary ($p'=0$) at the outlet. The centre of the passive flame is fixed at a quarter of the duct length, $b/L=0.25$. $\delta_f/L=0.1$, $\bar{T}_u=300$K, $\bar{T}_d=1200$K.}
	 \label{Fig:Effect_Ma}
	 \vspace*{00pt}
\end{figure}
\subsection{Effect of incoming mean flows and boundary conditions}
\label{sec:EigenFre_with_Ma}
Predictions are now carried out for the eigenvalue system in the presence of zero and non-zero incoming flow Mach numbers. With the inlet acoustically closed, both choked and open-end ($p'=0$) boundary conditions at the duct outlet are considered. 
As shown in Fig.~\ref{Fig:Effect_Ma}, both the frequencies and growth rates of the thermoacoustic modes vary with the flow Mach number. The modes for the open-end outlet boundary condition exhibit almost linear variation with the flow Mach number.

This is consistent with the propagation of the entropy wave significantly affecting the thermoacoustic modes for a system in which the boundary condition is dependent on the entropy perturbations (as is the case for a choked boundary). In neglecting the mean flow effect,  it is possible to introduce significant errors into theoretical predictions of thermoacoustic instabilities, even at relatively low flow Mach numbers. 

\section{Full-frequency prediction by combining the AE and the WKB solutions}
\label{sec:combination}
Dilution, forced cooling or wall heat transfer typically gives rise to mean temperature gradients on either side of the flame. Assuming that the flow properties in these regions vary slowly relative to the acoustic wavelengths, approximate high-frequency solutions can be applied to these regions. When the high-frequency condition is satisfied and the flow Mach number terms of order higher than $M^2$
can be neglected, the analytical modified WKB solutions proposed in our previous work \citep{Li_JSV_2017a} offer the potential for accurate predictions when the duct entropy disturbance can be neglected. The present AE method is superior in the low-frequency range, and apples in the presence of both acoustic and significant entropy waves. It is interesting to consider whether prediction over the full range of frequencies can be achieved by combining the low frequency AE solutions with the high frequency WKB solutions. 

\subsection{Pure cooling case with a linear temperature profile}
\label{sec:linear_cooling_case}
\begin{figure}[!ht]
	\centering
	\subfigure
  		{
\includegraphics[width=6cm]{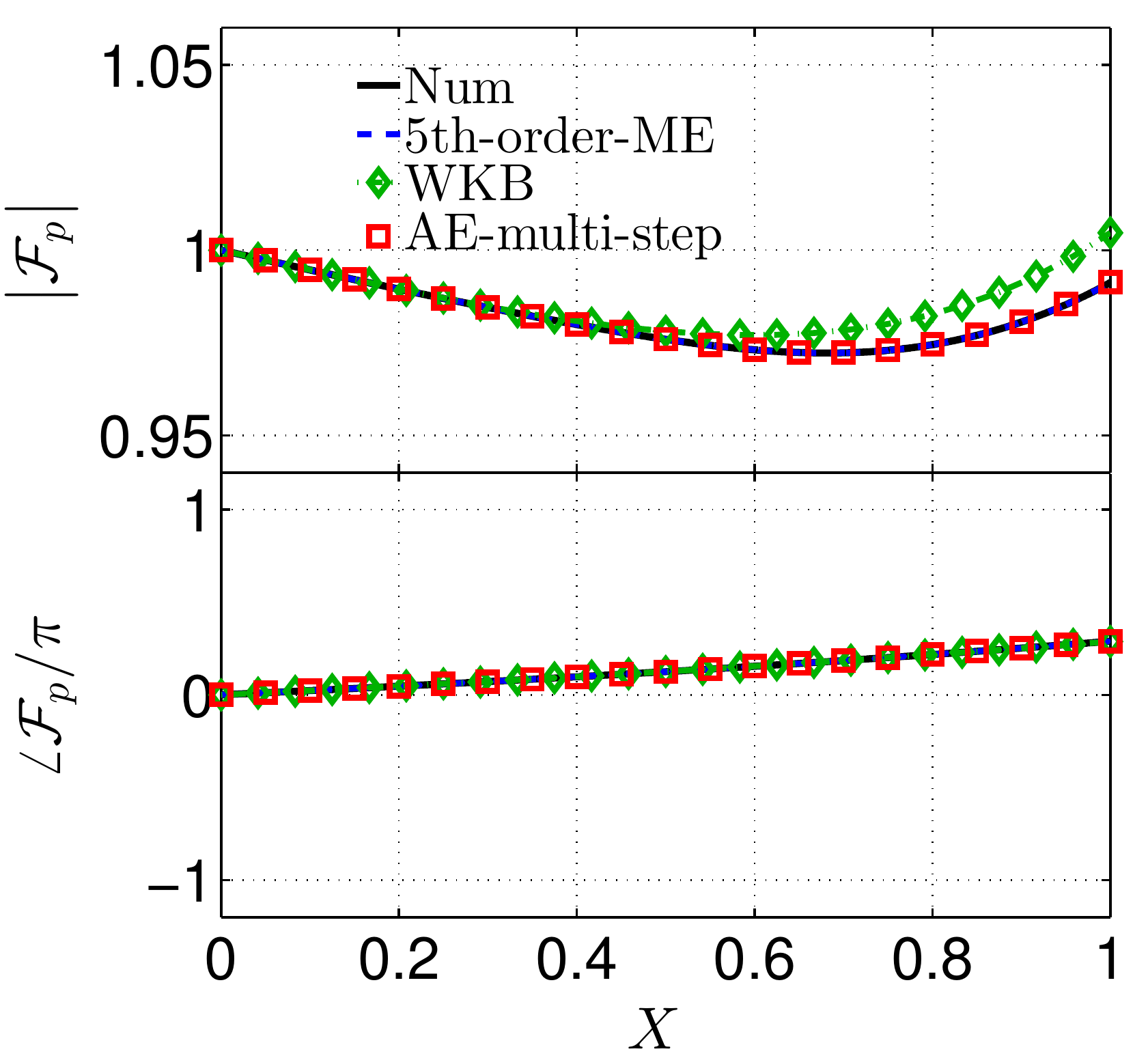}
  		}
  		\put (-190,150) {\normalsize$\displaystyle(a)$}
  		\vspace*{-0pt}
  		\hspace*{20pt}
  		\subfigure
  		  		{
\includegraphics[width=6cm]{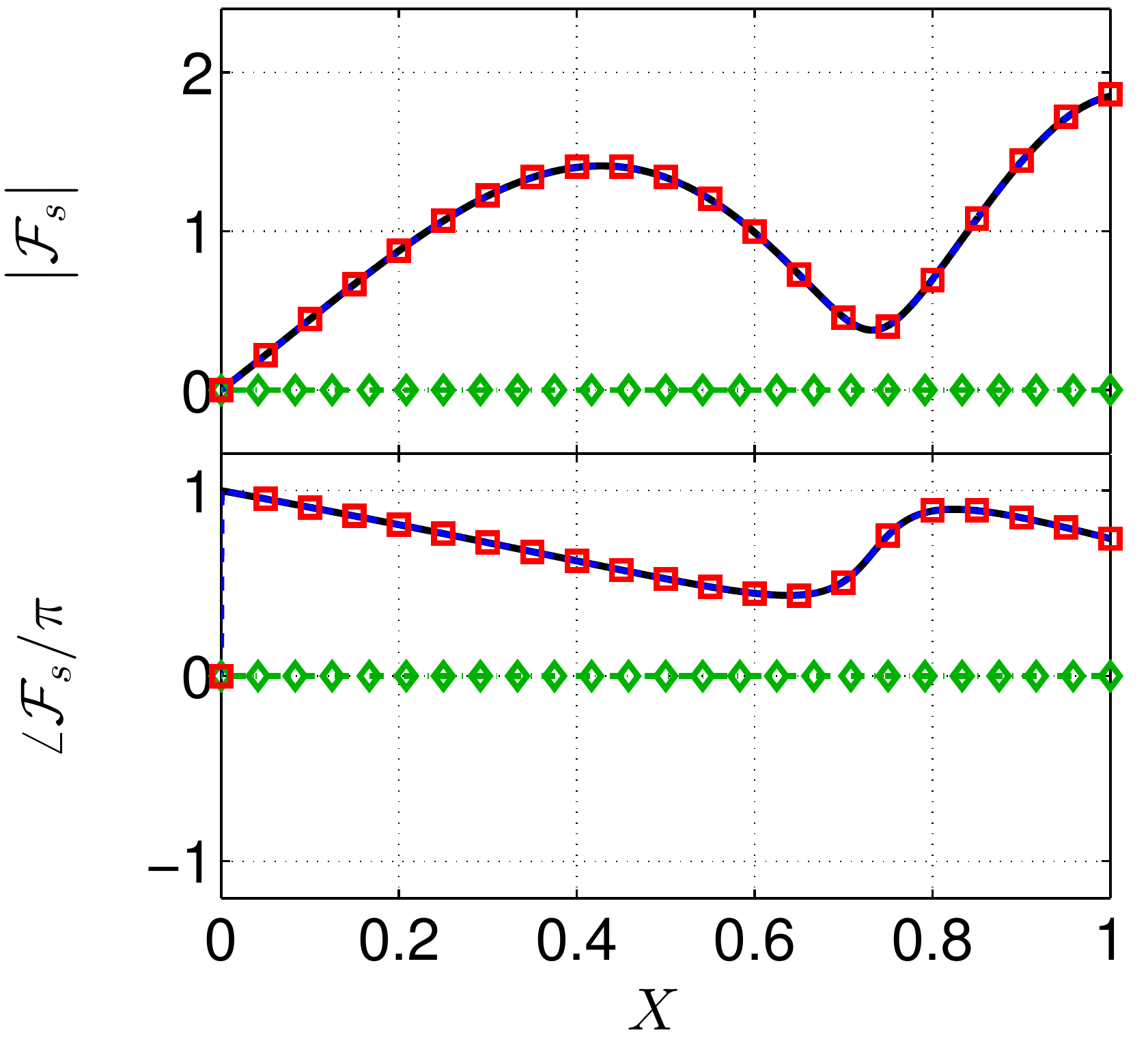}
  		}
  		\put (-190,150) {\normalsize$\displaystyle(b)$}
  		\vspace*{-0pt}

 	 \caption{Variations of (a) pressure $\mathcal{F}_p$ and (b) entropy $\mathcal{F}_s$ transfer functions as the function of the normalised axial position $X$ with no entropy disturbance ($\mathcal{F}_s|_{X=0}=0$) in the incoming mean flow: numerical LEEs simulation ($\mathbf{-}$), 5th-order ME method (\textcolor{blue}{$\mathbf{- -}$}), WKB method (\textcolor[rgb]{0,0.7,0}{$\diamondsuit$}) and multi-step AE method (\textcolor{red}{$\square$}) are presented. $\Omega=0.1$, $M_u=0.1$, $\bar{T}_u=1600$K, $\bar{T}_d=800$K.}
	 \label{Fig:TF_Cool_NonEn_f010_M010}
	 \vspace*{00pt}
\end{figure} 
\begin{figure}[!ht]
	\centering
	\subfigure
  		{
\includegraphics[width=6cm]{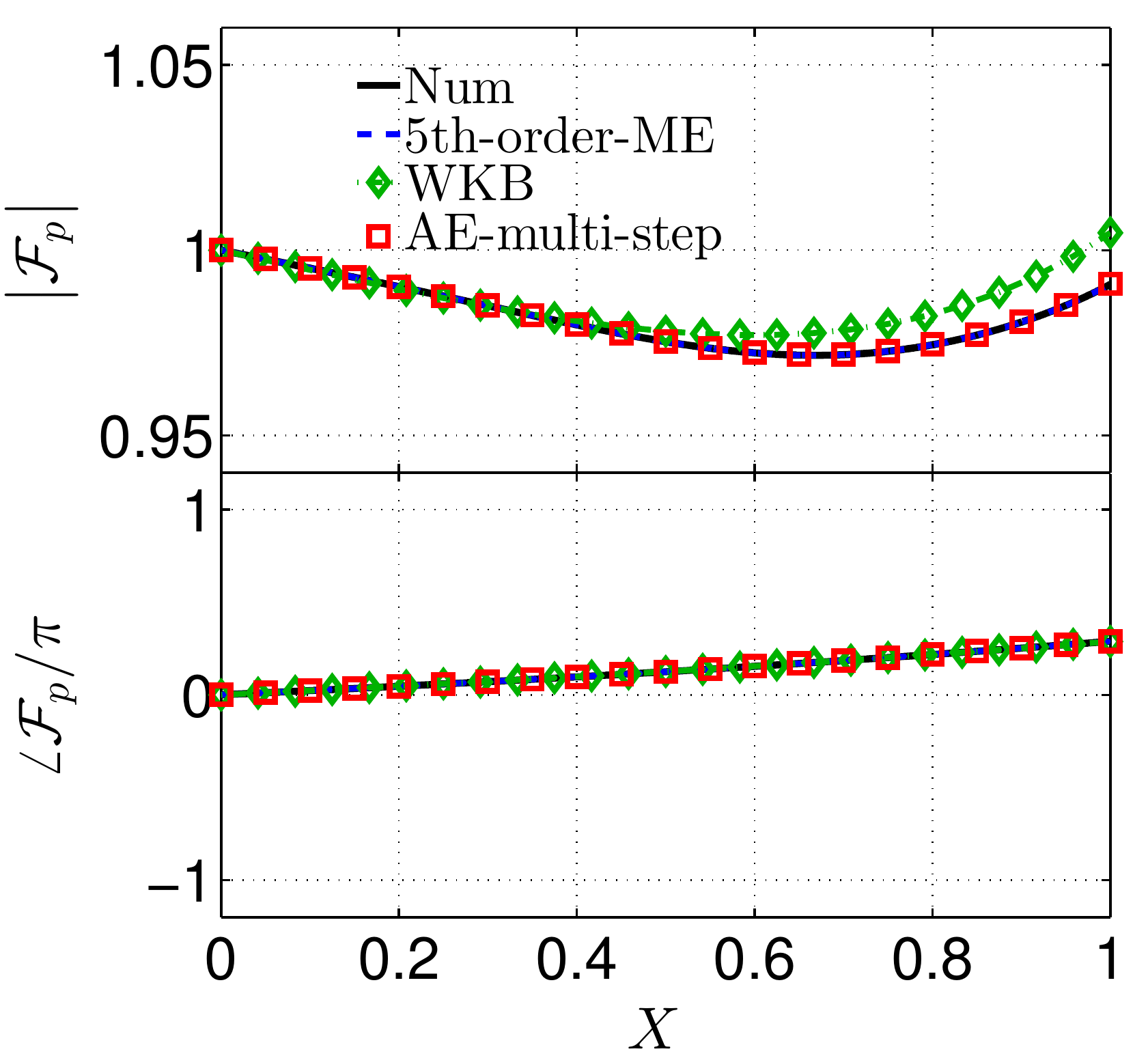}
  		}
  		\put (-190,150) {\normalsize$\displaystyle(a)$}
  		\vspace*{-0pt}
  		\hspace*{20pt}
  		\subfigure
  		  		{
\includegraphics[width=6cm]{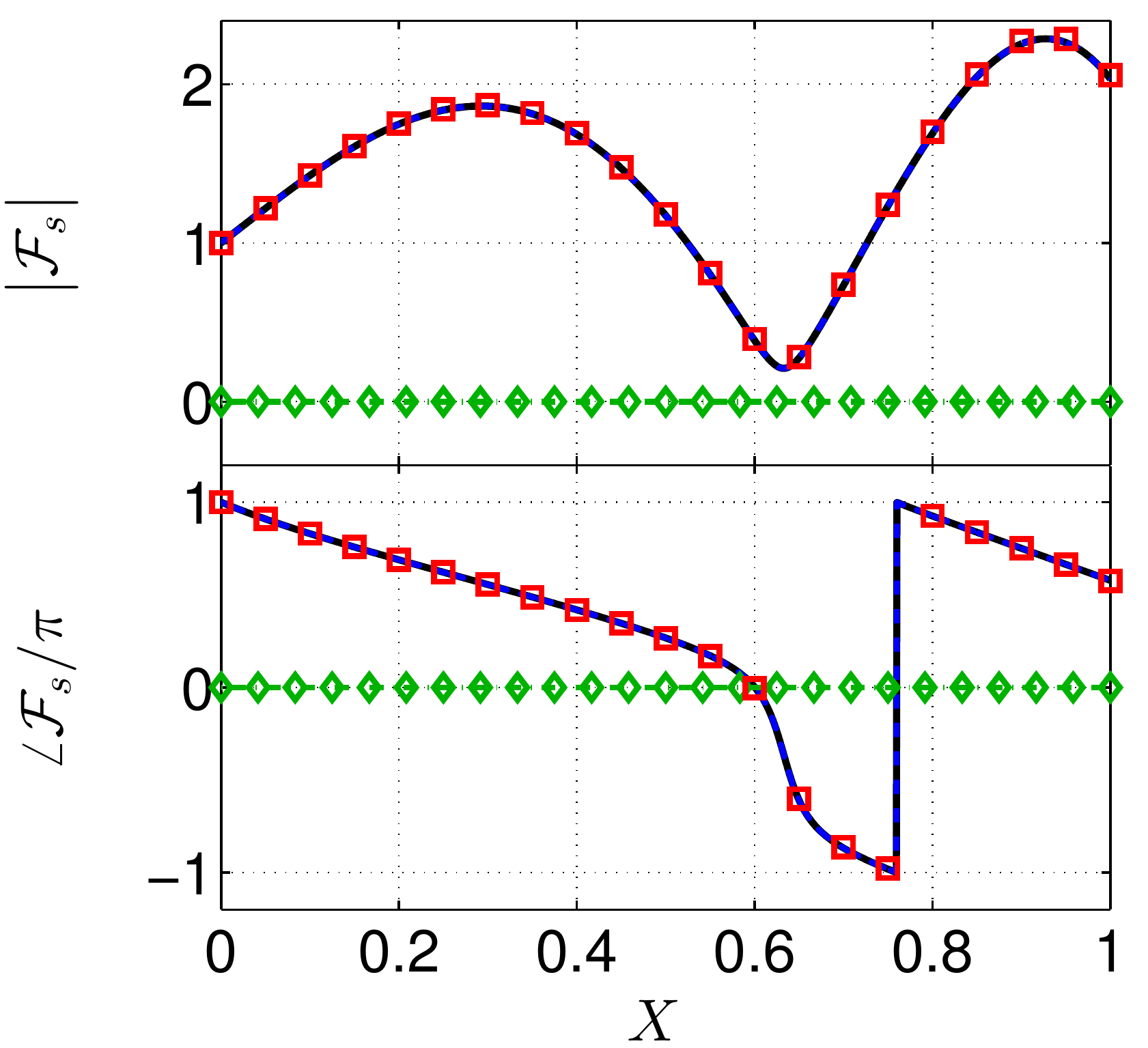}
  		}
  		\put (-190,150) {\normalsize$\displaystyle(b)$}
  		\vspace*{-0pt}

 	 \caption{Variations of (a) pressure $\mathcal{F}_p$ and (b) entropy $\mathcal{F}_s$ transfer functions as the function of the normalised axial position $X$ with non-zero entropy disturbance ($\mathcal{F}_s|_{X=0}=1/Z_u=-1$) in the incoming mean flow: numerical LEEs simulation ($\mathbf{-}$), 5th-order ME method (\textcolor{blue}{$\mathbf{- -}$}), WKB method (\textcolor[rgb]{0,0.7,0}{$\diamondsuit$}) and multi-step AE method (\textcolor{red}{$\square$}) are presented. $\Omega=0.1$, $M_u=0.1$, $\bar{T}_u=1600$K, $\bar{T}_d=800$K.}
	 \label{Fig:TF_Cool_with_En_f010_M010}
	 \vspace*{00pt}
\end{figure} 

The AE method is applied to the case of a $L=1\mathrm{m}$ long duct undergoing axial cooling from $1600$K to $800$K. Incident acoustic disturbances are at low normalised frequency $\Omega=0.1$. The acoustic transfer functions are still determined by the impedance of $Z_u=-1$ at $X=X_u=0$ as expressed in Eq.~\eqref{eq:BC_Xu}. The WKB solution derived in our previous work \citep{Li_JSV_2017a} is presented in  \ref{sec:WKB}. Figure~\ref{Fig:TF_Cool_NonEn_f010_M010} shows that the analytical AE solution agrees well with the numerical LEE and the ME results when it is assumed that there is no entropy perturbation upstream. The WKB method exhibits errors for the pressure transfer function at this low-frequency and does not provide any information on the entropy perturbation. 

For the same duct flow but with a non-zero entropy perturbation present in the incoming mean flow, Fig.~\ref{Fig:TF_Cool_with_En_f010_M010} shows that the proposed AE method still achieves accurate predictions for both the acoustic and entropy waves. 

\begin{figure}[!ht]
	\centering
	\subfigure
  		{
\includegraphics[width=6cm]{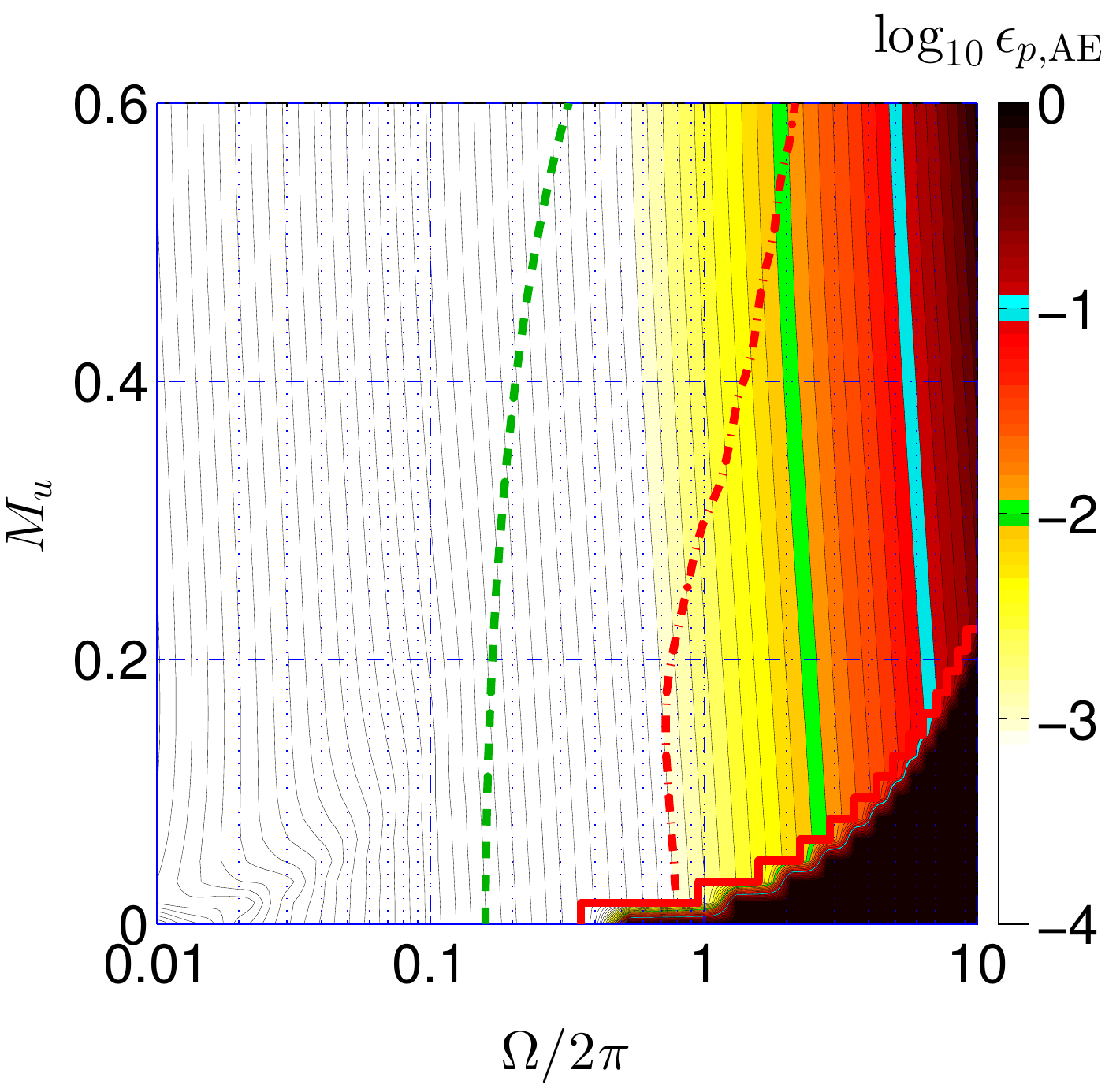}
  		}
  		\put (-190,150) {\normalsize$\displaystyle(a)$}
  		\vspace*{-0pt}
  		\hspace*{20pt}
  		\subfigure
  		  		{
\includegraphics[width=6cm]{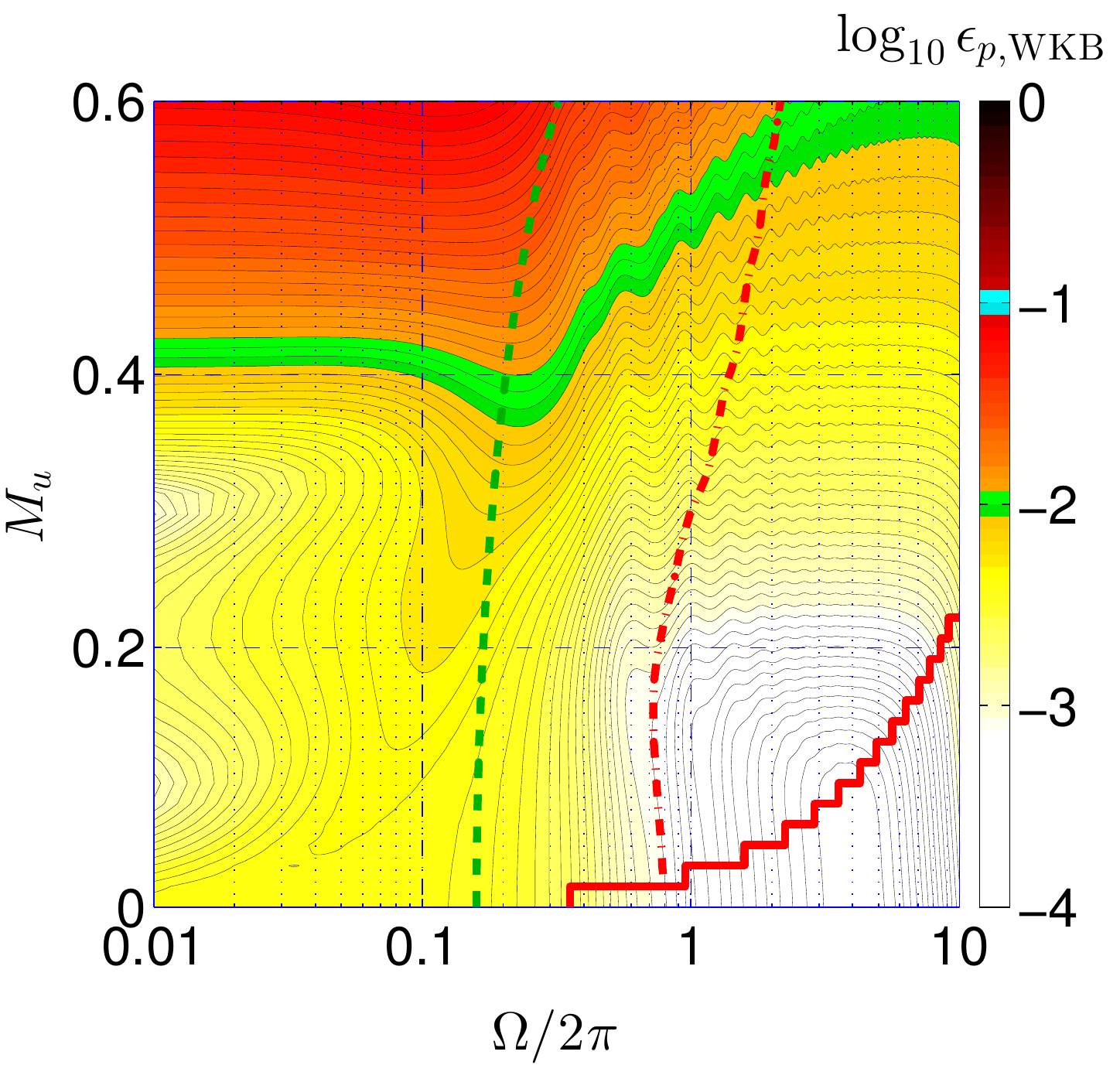}
  		}
  		\put (-190,150) {\normalsize$\displaystyle(b)$}
  		\vspace*{-0pt}

 	 \caption{Contour maps for the pressure error coefficients $\epsilon_p$ of (a) the AE solution and (b) the WKB solution: $\epsilon_{\mathrm{p,AE}}=\epsilon_{\mathrm{p,WKB}}$(\textcolor{red}{$\mathbf{-.}$}), $\Omega_{\mathrm{HF,WKB}}$(\textcolor[rgb]{0,0.7,0}{$\mathbf{- -}$}) and the upper frequency bound $\Omega_{\mathrm{max}}$ (\textcolor{red}{$\mathbf{-}$}) for the validity of the AE solutions are plotted on the contour, respectively. $N_s=5500$,  $\bar{T}_u=1600$K, $\bar{T}_d=800$K.}
	 \label{Fig:Contour_AE_WKB_a}
	 \vspace*{00pt}
\end{figure} 
\begin{figure}[!ht]
	\centering
\includegraphics[width=6cm]{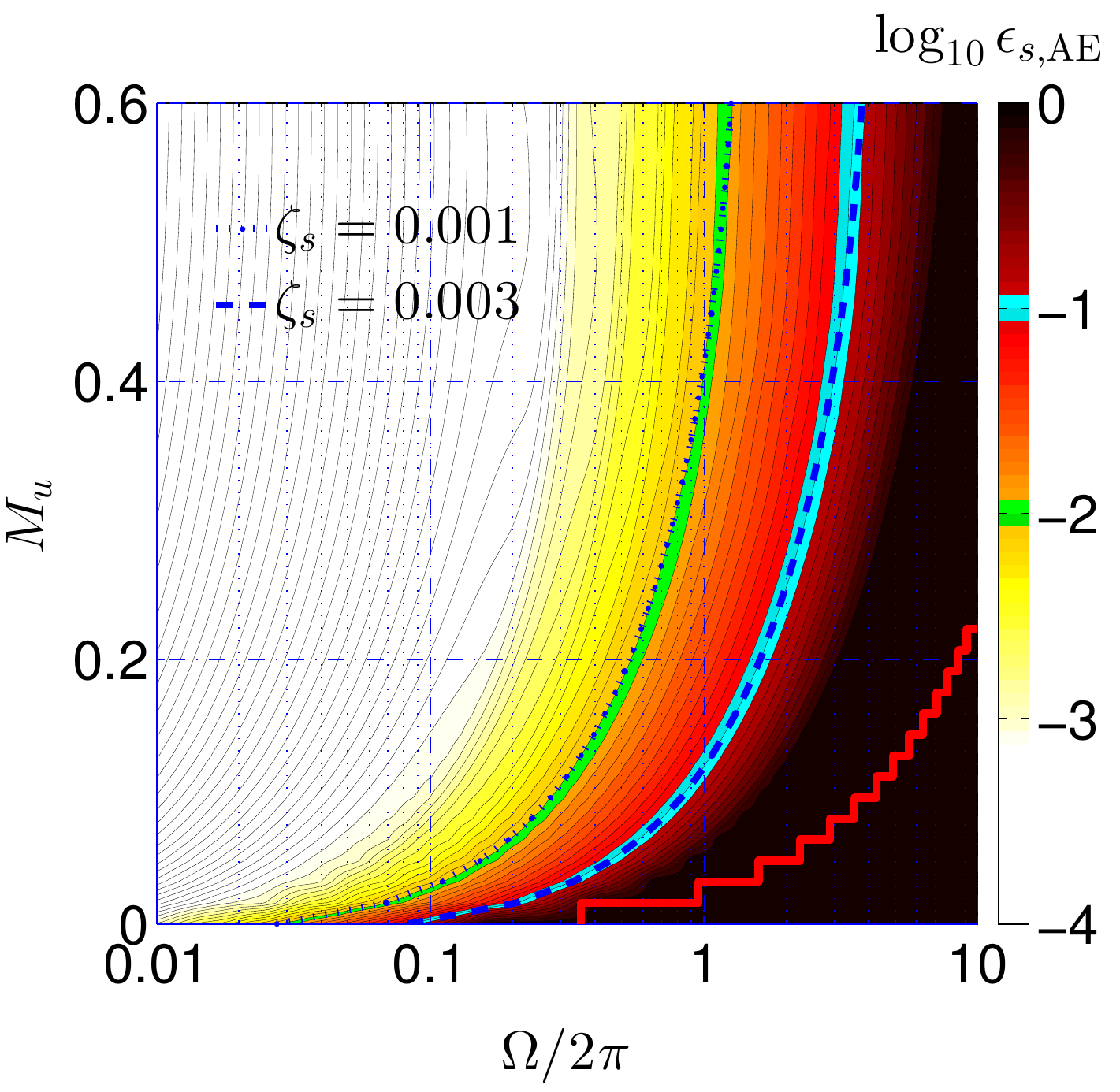}
  		\vspace*{-0pt}
 	 \caption{Contour map for entropic error coefficients $\epsilon_s$ of the AE solution: $N_s=5500$, $\bar{T}_u=1600$K, $\bar{T}_d=800$K.}
	 \label{Fig:Contour_AE_WKB_s}
	 \vspace*{00pt}
\end{figure}

\subsection{Range of the validity of the AE solutions}
\label{sec:AE_limit}
The range of the validity of the AE method is explored and compared to that of the modified WKB approximation method for the same cooling case as in Ref.~\citep{Li_JSV_2017a} (as is the same case in Sec.\ref{sec:linear_cooling_case}). The error coefficients of the WKB solution are similarly calculated by comparing to results from numerical simulation of the LEEs. The WKB solution and its corresponding high-frequency condition $\Omega_{\mathrm{HF}}$ can be found in the \ref{sec:WKB}. The error coefficient for the pressure is presented in Fig.~\ref{Fig:Contour_AE_WKB_a} as a function of the normalised frequency $\Omega$ and the incoming flow Mach number $M_u$. The AE solutions for pressure are accurate at low frequencies, but accuracy deteriorates at higher frequencies, especially in the lower right corner where the Mach number is also low. The entropy wavelength being much smaller than the distributed length means that more segments are required in the multi-step AE method.

Thankfully, there is an overlap region between the high-frequency condition $\Omega_{\mathrm{HF}}$ for the WKB solution and the equal-accuracy line of the two methods. The modified WKB approximation method can obtain accurate results for acoustic waves in the high frequency region when the flow Mach number varies from low and moderate to even high, e.g. $M=0.6$. The AE solution is able to provide accurate predictions for the entropy wave across the range of low frequencies and even very low-speed flows in which the entropy disturbance plays a crucial role in the predictions of thermoacoustic modes, as shown in Fig.~\ref{Fig:Contour_AE_WKB_s}.
Hence a combination of the proposed AE method for low frequencies and the modified WKB approximation method for high frequencies can achieve a full-frequency prediction for thermoacoustic system.

%
\section{Conclusions}
\label{sec:conclusion}
Analytical solutions for the linearised Euler equations governing the transport of planar 1-D disturbances in duct flows sustaining distributed heat transfer have been derived. Asymptotic expansion solutions are obtained by performing an asymptotic analysis for low frequencies; this allows the reconstruction of acoustic and entropy wave propagating through the region. The linearised Euler equations (LEEs) are first rewritten as ordinary differential equations (ODEs) with varying coefficients via flow invariants defined by the perturbation conservation of mass, momentum and energy. As a consequence, direct analytical expressions for the acoustic and entropy waves can be derived via an asymptotic expansion with second-order terms in frequency neglected. A multi-step strategy can be applied to extend the AE solutions to higher frequencies and smaller Mach numbers in order to capture the much smaller wavelength of the entropy perturbations at these conditions. It has been shown that an eigenvalue system can be built for ducts containing a region of distributed heat communication; this can be used to predict the thermoacoustic modes of the duct for given acoustic boundary conditions.

Validation of the analytical solutions has been performed for duct regions undergoing both heating and cooling. Predictions from the proposed AE method are in good agreement with numerical simulation of the LEEs, when the low-frequency condition $\Omega\gg\Omega_{\mathrm{LF}}$ is satisfied. It is shown that compact models, which do not account for the spatial variation in the heat transfer, incorrectly predict the eigenfrequencies and growth rates, even for very thin heat transfer regions. Comparisons of the mode structures reveal that different entropy wave strengths are generated by the acoustic wave interacting with the steady heat source. The downstream entropy perturbations convected by the mean flow then impact the acoustic field upon encountering a density- or entropy-dependent downstream boundary. The assumption of no mean flow is seen to cause incorrect prediction of the thermoacoustic modes because this acoustic-entropy coupling is neglected. Therefore, the AE solutions have been validated as being able to correct the wrong predictions of the compact model and accurately reconstruct the acoustic and entropy waves downstream of the flow variation region. It has further been demonstrated that the generation and propagation of the entropy wave have to be taken into account in order to accurately predict the thermoacoustic modes in the low-frequency domain, especially when density- or entropy-dependent boundary conditions are found at the duct ends.

For a duct undergoing steady cooling, it is shown that a frequency overlap region occurs between the low-frequency condition of the AE method and the high-frequency condition of complementary models provided by the modified WKB approximation method. Moreover, the modified WKB solution can extend the application of theoretical analysis to a full-frequency range from low to moderate even to large subsonic Mach numbers, e.g. $M=0.6$. Therefore, a prediction of thermoacoustic modes can be achieved in a full-frequency range and a very wide subsonic Mach number range by combining the AE solutions for low frequencies with the WKB solution for high frequencies.

\section*{Acknowledgement}
The authors would like to gratefully acknowledge financial support from the Chinese National Natural Science Funds for National Natural Science Foundation of China (Grant no. 51806006, 11927802, and U1837211), and National Major Science and Technology Projects of China (2017-III-0004-0028). The European Research Council (ERC) Consolidator grant AFIRMATIVE (2018--2023) is also gratefully acknowledged.


\appendix
\section{Derivation of the eigenvalue system using the AE solutions}
\label{sec:eigenvalue}
The acoustic wave is assumed to be a superposition of the plane waves propagating upstream and downstream. With uniform temperature distributions and mean flow, the wavenumbers of the acoustic and entropy perturbations are expressed as:
\begin{equation}
k^+ = - \frac{\omega/\bar{c}}{1+M}, \quad k^- = \frac{\omega/\bar{c}}{1-M} , \quad k_0 = - \frac{\omega}{\bar{u}}.
\end{equation}

The wavenumbers are further reduced by the normalised variables in Eq.~\eqref{eq:reduced_variables}, expressed as:
\begin{equation}
K^+ = k^+ L = - \frac{\Omega / \bar{\boldsymbol{c}}}{1 + M}, \quad
K^- = k^- L =  \frac{\Omega / \bar{\boldsymbol{c}}}{ 1- M}, \quad 
K_0 = k_0 L =  -\frac{\Omega}{\bar{\boldsymbol{U}}}
\end{equation}

Analytical expressions for normalised acoustic perturbations are thus given by:
\begin{align}
\label{eq:Sol_p_mean_T}
\hat{\boldsymbol{P}} = A^+ e^{\mathrm{i} K^+ X} +A^- e^{\mathrm{i} K^- X}, \\
\label{eq:Sol_rho_mean_T}
\hat{\boldsymbol{\rho}} = \frac{1}{\bar{\boldsymbol{c}}^2} \left( A^+ e^{\mathrm{i} K^+ X} +A^- e^{\mathrm{i} K^- X}  - E e^{\mathrm{i} K_0 X} \right),\\
\label{eq:Sol_u_mean_T}
\hat{\boldsymbol{U}} = \frac{1}{\bar{\boldsymbol{\rho}} \bar{\boldsymbol{c}}} \left( A^+ e^{\mathrm{i} K^+ X} -A^- e^{\mathrm{i} K^- X} \right),
\end{align}
where the acoustic and entropy wave amplitudes $\mathcal{A}^+, \mathcal{A}^-, \mathcal{E}$ are normalised by the constant upstream flow properties,
\begin{equation}
A^+ = \frac{\mathcal{A}^+}{\bar{\rho}_u \bar{c}_u^2}, \quad
A^- = \frac{\mathcal{A}^-}{\bar{\rho}_u \bar{c}_u^2}, \quad 
E = \frac{\mathcal{E}}{\bar{\rho}_u \bar{c}_u^2}.
\end{equation}

The acoustic perturbations can be recovered from the wave amplitude vector via $\mathcal{M}^{w2p}$, expressed as:
\begin{equation}
\begin{bmatrix}
\hat{\boldsymbol{P}} \vspace{0.2cm} \\
\hat{\boldsymbol{\rho}} \vspace{0.2cm} \\
\hat{\boldsymbol{U}}\\
\end{bmatrix}
=
 \mathcal{M}^{w2p} 
\begin{bmatrix}
A^+ \vspace{0.2cm} \\
A^- \vspace{0.2cm} \\
E\\
\end{bmatrix}
=
\begin{bmatrix}
1 &1 &0 \vspace{0.2cm} \\
\dfrac{1}{\bar{\boldsymbol{c}}^2} &\dfrac{1}{\bar{\boldsymbol{c}}^2} &-\dfrac{1}{\bar{\boldsymbol{c}}^2}\vspace{0.2cm} \\
\dfrac{1}{\bar{\boldsymbol{\rho}} \bar{\boldsymbol{c}}} &-\dfrac{1}{\bar{\boldsymbol{\rho}} \bar{\boldsymbol{c}}} &0\\
\end{bmatrix}
\begin{bmatrix}
e^{\mathrm{i} K^+ X} & & \vspace{0.2cm} \\
 &e^{\mathrm{i} K^- X}  & \vspace{0.2cm} \\
 &  &e^{\mathrm{i} K_0 X} \\
\end{bmatrix}
\begin{bmatrix}
A^+ \vspace{0.2cm} \\
A^- \vspace{0.2cm} \\
E\\
\end{bmatrix}\,.
\end{equation}

Assuming no entropy perturbation ($E_u=0$) is present upstream of the distributed steady heat source, one gets the transfer matrices of the acoustic perturbations on both sides of the distributed region, respectively, given by:
\begin{equation}
\begin{bmatrix}
\hat{\boldsymbol{P}} \vspace{0.2cm} \\
\hat{\boldsymbol{\rho}} \vspace{0.2cm} \\
\hat{\boldsymbol{U}}\\
\end{bmatrix}_u
=
\mathcal{M}^{w2p}_u
\begin{bmatrix}
A_u^+ \vspace{0.2cm} \\
A_u^- \\
\end{bmatrix}
=
\begin{bmatrix}
1 &1  \vspace{0.2cm} \\
\dfrac{1}{\bar{\boldsymbol{c}}_u^2} &\dfrac{1}{\bar{\boldsymbol{c}}_u^2} \vspace{0.2cm} \\
\dfrac{1}{\bar{\boldsymbol{\rho}}_u \bar{\boldsymbol{c}}_u} &-\dfrac{1}{\bar{\boldsymbol{\rho}}_u \bar{\boldsymbol{c}}_u} \\
\end{bmatrix}
\begin{bmatrix}
e^{\mathrm{i} K_u^+ X_u} & \vspace{0.2cm} \\
 &e^{\mathrm{i} K_u^- X_u}  \\
\end{bmatrix}
\begin{bmatrix}
A_u^+ \vspace{0.2cm} \\
A_u^- \\
\end{bmatrix}\,,
\end{equation}

\begin{equation}
\begin{bmatrix}
\hat{\boldsymbol{P}} \vspace{0.2cm} \\
\hat{\boldsymbol{\rho}} \vspace{0.2cm} \\
\hat{\boldsymbol{U}}\\
\end{bmatrix}_d
=
\mathcal{M}^{w2p}_d
\begin{bmatrix}
A_d^+ \vspace{0.2cm} \\
A_d^- \vspace{0.2cm} \\
E_d\\
\end{bmatrix}
=
\begin{bmatrix}
1 &1 &0 \vspace{0.2cm} \\
\dfrac{1}{\bar{\boldsymbol{c}}_d^2} &\dfrac{1}{\bar{\boldsymbol{c}}_d^2} &-\dfrac{1}{\bar{\boldsymbol{c}}_d^2}\vspace{0.2cm} \\
\dfrac{1}{\bar{\boldsymbol{\rho}}_d \bar{\boldsymbol{c}}_d} &-\dfrac{1}{\bar{\boldsymbol{\rho}}_d \bar{\boldsymbol{c}}_d} &0\\
\end{bmatrix}
%
%
\begin{bmatrix}
A_d^+ \vspace{0.2cm} \\
A_d^- \vspace{0.2cm} \\
E_d\\
\end{bmatrix}\,.
\end{equation}

Then the acoustic transfer matrix of the AE method is used to deal with the distributed region to connect the upstream and downstream regions, written as:
\begin{equation}
\left[\boldsymbol{\mathrm{T}}\right]_{u}^{d}
\mathcal{M}^{w2p}_u
\begin{bmatrix}
A_u^+ \vspace{0.2cm} \\
A_u^-  \\
\end{bmatrix}
-
\mathcal{M}^{w2p}_d
\begin{bmatrix}
A_d^+ \vspace{0.2cm} \\
A_d^- \vspace{0.2cm} \\
E_d\\
\end{bmatrix}
=0\,.
\end{equation}

In the form of wave amplitudes, the acoustically closed boundary condition of the duct inlet is expressed as:
\begin{equation}
\label{eq:A_LBC}
\hat{\boldsymbol{U}}( {X=0})
=
\begin{bmatrix}
\dfrac{1}{\bar{\boldsymbol{\rho}}_u \bar{\boldsymbol{c}}_u} &-\dfrac{1}{\bar{\boldsymbol{\rho}}_u \bar{\boldsymbol{c}}_u} \\
\end{bmatrix}
\begin{bmatrix}
A_u^+ \vspace{0.2cm} \\
A_u^- \\
\end{bmatrix}=0\,,
\end{equation}
and the choked boundary condition of the outlet, $- \hat{\boldsymbol{P}}/\bar{\boldsymbol{P}} +\hat{\boldsymbol{\rho}}/\bar{\boldsymbol{\rho}} +  2 \hat{\boldsymbol{U}}/ \bar{\boldsymbol{U}}\, |_{X=1}=0$, is given by:
\begin{equation}
\label{eq:A_choked BC}
\begin{bmatrix}
-1/\bar{\boldsymbol{P}}_d &1/\bar{\boldsymbol{\rho}}_d &2/\bar{\boldsymbol{U}}_d 
\end{bmatrix}
\begin{bmatrix}
1 &1 &0 \vspace{0.2cm} \\
\dfrac{1}{\bar{\boldsymbol{c}}_d^2} &\dfrac{1}{\bar{\boldsymbol{c}}_d^2} &-\dfrac{1}{\bar{\boldsymbol{c}}_d^2}\vspace{0.2cm} \\
\dfrac{1}{\bar{\boldsymbol{\rho}}_d \bar{\boldsymbol{c}}_d} &-\dfrac{1}{\bar{\boldsymbol{\rho}}_d \bar{\boldsymbol{c}}_d} &0\\
\end{bmatrix}
%
%
\begin{bmatrix}
e^{\mathrm{i} K_d^+ L_d} & & \vspace{0.2cm} \\
 &e^{\mathrm{i} K_d^- L_d}  & \vspace{0.2cm} \\
 &  &e^{\mathrm{i} K_{0,d} L_d} \\
\end{bmatrix}
\begin{bmatrix}
A_d^+ \vspace{0.2cm} \\
A_d^- \vspace{0.2cm} \\
E_d\\
\end{bmatrix}
=0
\end{equation}
where $L_d=1-X_d$ is the normalised length of the downstream region.

The formula is rewritten using three coefficients $R_1$, $R_2$ and $R_3$ as:
\begin{equation}
R_1  A_d^+ + R_2 A_d^- +R_3 E_d=0\,,
\end{equation}
where 
\begin{equation}
R_1 = \left( -\frac{1}{\bar{\boldsymbol{P}}_d }+ \frac{1}{\bar{\boldsymbol{\rho}}_d \bar{\boldsymbol{c}}_d^2} + \frac{2}{\bar{\boldsymbol{\rho}}_d \bar{\boldsymbol{c}}_d \bar{\boldsymbol{U}}_d }\right)e^{\mathrm{i} K_d^+ L_d}\,,\quad
R_2 = \left( -\frac{1}{\bar{\boldsymbol{P}}_d } + \frac{1}{\bar{\boldsymbol{\rho}}_d \bar{\boldsymbol{c}}_d^2} - \frac{2}{\bar{\boldsymbol{\rho}}_d \bar{\boldsymbol{c}}_d \bar{\boldsymbol{U}}_d } \right)e^{\mathrm{i} K_d^- L_d}\,, \quad
R_3 =- \frac{1}{\bar{\boldsymbol{\rho}}_d \bar{\boldsymbol{c}}_d^2}e^{\mathrm{i} K_{0,d} L_d} \,.
\end{equation}

It should be noted that the choked boundary condition turns into an acoustically closed condition, $\hat{\boldsymbol{U}}( {X=1}) =0$ if there is no mean flow in the duct. Then the coefficients are expressed as:
\begin{equation}
\label{eq:A_RBC_noflow}
R_1 =  \frac{1}{\bar{\boldsymbol{\rho}}_d \bar{\boldsymbol{c}}_d }e^{\mathrm{i} K_d^+ L_d}\,,\quad
R_2 =  - \frac{1}{\bar{\boldsymbol{\rho}}_d \bar{\boldsymbol{c}}_d} e^{\mathrm{i} K_d^- L_d}\,, \quad
R_3 =0 \,.
\end{equation}

Finally, the eigenvalue system using the acoustic transfer matrix of the AE method is derived:
\begin{equation}
\label{eq:AE_Eigen_Appendix}
\mathbb{M}^{\mathrm{AE}} \cdot C^{\mathrm{AE}} =
\left[
\begin{array}{c|c}
\begin{matrix}
   \vspace{0.2cm}\\
 \left[\boldsymbol{\mathrm{T}}\right]_{u}^{d}
\mathcal{M}^{w2p}_u \vspace{0.2cm}\\
 \\
 \end{matrix} 
 &
 \begin{matrix}
 -\mathcal{M}^{w2p}_d 
 \end{matrix} \\ \hline
 \begin{matrix}
   &  \vspace{-0.2cm}\\
  1/\bar{\boldsymbol{\rho}}_u \bar{\boldsymbol{c}}_u
&-1/\bar{\boldsymbol{\rho}}_u \bar{\boldsymbol{c}}_u \vspace{0.2cm}\\
 0 & 0\\
 \end{matrix}
 &\begin{matrix}
   & & \vspace{-0.2cm}\\
 0 &0 &0 \vspace{0.2cm}\\
  R_1 &R_2 &R_3 \\
 \end{matrix} 
\end{array}
\right]_{5\times 5}
 \begin{bmatrix}
A_u^+ \vspace{0.2cm}\\
A_u^- \vspace{0.2cm}\\
A_d^+ \vspace{0.2cm}\\
A_d^- \vspace{0.2cm}\\
 E_d\\
\end{bmatrix}
 = \mathbf{0}\,.
\end{equation}

\section{The WKB solution and the high-frequency condition}
\label{sec:WKB}
The WKB solution and the high-frequency assumption are exactly reproduced from our previous work \citep{Li_JSV_2017a}, except the upstream parameters here are denoted by the subscript `$u$' instead of `$1$'. Readers can refer to our previous work for detailed derivations.

Assuming that the entropy wave has negligible effect on the acoustic field in the high-frequency domain, the linearised Euler equations (LEEs) are formulated to obtain an acoustic wave equation in the form of the pressure perturbation, written as:
\begin{equation}
\label{eq:wave_eq_final}
\begin{split}
&\left(1 - M^2 + \mathrm{i}\frac{2 M^2}{k_0} \frac{\mathrm{d} M}{\mathrm{d} x}\right)\frac{\mathrm{d}^2 \hat{p}}{\mathrm{d}x^2}
 -
\left(\big(1 - (3 + \gamma) M^2\big)\alpha + \mathrm{i} 2 M k_0 + \mathrm{i} \frac{M \beta}{k_0} - \mathrm{i}\frac{2 M \alpha^2}{k_0}\right)\frac{\mathrm{d} \hat{p}}{\mathrm{d}x} \\
 +
 &\left(k_0^2 + \mathrm{i} (2 + \gamma)M k_0 \alpha - \mathrm{i} 2\gamma k_0 M^2 \frac{\mathrm{d}M}{\mathrm{d}x} + (2 - \gamma)M^2 \beta + (4 \gamma - 5)M^2 \alpha^2 \right) \hat{p} = 0
 \end{split}
\end{equation}
where 
\begin{equation}
k_0=\omega/\bar{c}, \quad M=\bar{u}/\bar{c}, \quad \alpha = \frac{1}{\bar{\rho}}\frac{\mathrm{d} \bar{\rho}}{\mathrm{d}x}, \quad\beta = \frac{1}{\bar{\rho}}\frac{\mathrm{d}^2 \bar{\rho}}{\mathrm{d}x^2}\,.
\end{equation}

With the high-frequency condition $|k_0|\gg|\alpha|$ satisfied, the analytical solution for the pressure perturbation is derived by the modified WKB approximation method, expressed in the form of separate amplitude $a(x)$ and phase factors $b(x)$ as:
\begin{equation}
\hat{p}=\mathcal{A} \exp\left( \int_{x_u}^{x} a(x)+\mathrm{i} b(x) ~\mathrm{d} x \right)
\end{equation}
where the Mach number terms of order higher than $M^2$ in the amplitude factor $a(x)$ are assumed negligible. 

The solution of pressure perturbation is obtained by substituting the modified WKB solution into the wave equation (Eq.~\eqref{eq:wave_eq_final}), given by
\begin{equation}
\label{eq:p_final}
\hat{p}(x, \omega) = \mathcal{A}^+ \mathcal{P}^+(x, \omega) + \mathcal{A}^- \mathcal{P}^-(x, \omega)
\end{equation}
where 
\begin{equation}
\label{eq:p+}
\mathcal{P}^+(x, \omega) = \left(\frac{\bar{\rho}}{\bar{\rho}_u}\right)^{1/4}\frac{1 + M_u}{1 + M} \frac{\exp\left(\gamma M_u - \frac{\gamma}{4} M_u^2 - \frac{\gamma^2 - 1}{3} M_u^3\right)}{\exp\left(\gamma M - \frac{\gamma}{4} M^2 - \frac{\gamma^2 - 1}{3} M^3\right)} \exp\left(-\mathrm{i}\omega \int_{x_u}^{x} \frac{\mathrm{d}x}{\bar{c} + \bar{u}} \right)\,,
\end{equation}
\begin{equation}
\label{eq:p-}
\mathcal{P}^-(x, \omega) = \left(\frac{\bar{\rho}}{\bar{\rho}_u}\right)^{1/4}\frac{1 - M_u}{1 - M} \frac{\exp\left(\gamma M + \frac{\gamma}{4} M^2 - \frac{\gamma^2 - 1}{3} M^3\right)}{\exp\left(\gamma M_u + \frac{\gamma}{4} M_u^2 - \frac{\gamma^2 - 1}{3} M_u^3\right)}\exp\left(\mathrm{i}\omega  \int_{x_u}^{x} \frac{\mathrm{d}x}{\bar{c} - \bar{u}} \right)\,,
\end{equation}
and constant coefficients $\mathcal{A}^+$ and $\mathcal{A}^+$ can be determined by the initial values or the boundary conditions.
The high-frequency condition that has to be satisfied for the WKB solution is expressed as:
\begin{equation}
\Omega \gg \Omega_{\mathrm{HF}} = \frac{L}{\bar{c}_u} \max\Big| \frac{\bar{c}}{1-\gamma M^2}  \frac{1}{\bar{T}} \frac{\mathrm{d} \bar{T}}{\mathrm{d} x}\Big|\,.
\end{equation}

\bibliographystyle{elsarticle-num}
\bibliographystyle{unsrt}
%

%
%
%

\end{document}